%% file: online-hpcs18.tex
\documentclass[12pt]{article}
\usepackage{booktabs} 
\usepackage{multicol}
\usepackage{multirow}
\usepackage[ruled,vlined]{algorithm2e}
\usepackage{graphicx}
\usepackage{float}
\usepackage{dblfloatfix}
\usepackage{xcolor}
\usepackage{amsmath}
\usepackage{url}

\usepackage[noend]{algpseudocode}
\usepackage[labelsep=quad,indention=10pt]{subfig}
\usepackage[bottom]{footmisc}
\usepackage{ tablefootnote}
\usepackage{lscape}
\usepackage{adjustbox}

\definecolor{ForestGreen}{RGB}{63,142,38}
\newcommand{\ali}[1]{{\color{black}#1}}
\newcommand{\alia}[1]{{\color{black}#1}}

\newcommand{\flo}[1]{{\color{black}#1}}

\newcommand{\simdag}{\mbox{SG-SD}}
\newcommand{\msg}{\mbox{SG-MSG}}

\begin{document}
\title{Performance Reproduction and Prediction of \\ Selected Dynamic Loop Scheduling Experiments} 

%

\author{Ali Mohammed, Ahmed Eleliemy, and Florina M. Ciorba\\
	Department of Mathematics and Computer Science\\
		University of Basel, Switzerland}

\date{}
\maketitle
\clearpage
\tableofcontents
\clearpage

\input{0.tex}
\input{1.tex}

\input{2.tex}
\input{3.tex}
\input{4.tex}

\input{5.tex}

\input{6.tex}
\input{7.tex}

\input{8.tex}
\input{9.tex}

\bibliographystyle{ieeetr}
\bibliography{citedatabase} 

\end{document}

%% file: 0.tex
\begin{abstract}
\label{sec:abs}
%
%
Scientific applications are complex, large, and often exhibit irregular and stochastic behavior.
%
The use of efficient loop scheduling techniques, from static to \ali{fully} dynamic, in computationally-intensive applications characterized by large data-parallel loops\ali{,} is crucial for improving their performance, often degraded by load imbalance, on high-performance computing~(HPC) platforms.
%
%
%
A number of dynamic loop scheduling~(DLS) techniques has been proposed between \ali{the late} 1980's and early 2000's and efficiently used in scientific applications.
In most cases, the computing systems on which they have been tested and validated are no longer available. 
The use of DLS for the purpose of improving the performance of computationally-intensive scientific applications executing on modern HPC platforms is of increased significance today as system-induced load imbalance is exacerbated due to systems diversity, complexity, increased size, increased heterogeneity, and massively parallel nature. 
\alia{This work is concerned with the minimization of the sources of uncertainty in the implementation of DLS techniques to avoid unnecessary influences on the performance of scientific applications.}
\alia{Therefore,} it is important to ensure that the DLS techniques employed in scientific applications today adhere to their original design goals and specifications. 
%
%
The goal of this work is to \alia{ attain and increase the trust in the implementation of DLS techniques in today’s studies. To achieve this goal,} the performance of a selection of scheduling experiments from the 1992 original work that introduced \emph{factoring}, an efficient DLS technique proposed for shared-memory systems, \alia{is reproduced and predicted both}, via simulative and native experimentation.
\alia{The scientific challenge is the reproduction of the performance of the past experiments with incomplete information, such as the computing system characteristics and the implementation details.}
%
%
%
%
The selected scheduling experiments involve two computational kernels and four loop scheduling techniques. 
%
%
The experiments show that the simulation reproduces the performance achieved on the past computing platform and accurately predicts the performance achieved on the present computing platform. 
The performance reproduction and prediction confirms that the present implementation of these DLS techniques \alia{considered} both, in simulation and natively, adheres to their original description.
Moreover, the simulative and native experiments follow the expected performance behavior for the considered scheduling scenarios.
\alia{The results confirm the hypothesis that reproducing experiments of identical scheduling scenarios on past and modern hardware leads to an entirely different behavior from expected.}
%
This work paves the way towards additional simulative and native experimentation using further DLS techniques in the future. 
\end{abstract}


%% file: 1.tex
\section{Introduction}  
\label{sec:introduction}
\flo{Dynamic loop scheduling~(DLS) is an effective scheduling approach employed in computationally-intensive scientific applications for the purpose of optimizing their performance in the presence of load imbalance caused by problem, algorithmic, and systemic characteristics.
The DLS techniques dynamically schedule the work contained in the parallel loop iterations among the parallel processing units whenever they become available and request work.}
\flo{Over the years, DLS techniques have successfully been used in 
scientific applications, such as, N-body simulations, computational fluid dynamics, radar signal processing~\cite{AWFBC}, and computer vision application~\cite{EPSIA}. 
}

One of the well-known and efficient DLS techniques is \emph{factoring}, introduced by Hummel et al.~\cite{FAC} in 1992. 
Therein, the performance of the IBM Research Parallel Processor Prototype \flo{(hereafter, the RP3)} system~\cite{RP3} \flo{was} compared for the execution of three computational kernels: matrix multiplication, adjoint convolution, and Gauss-Jordan elimination, using four scheduling techniques:~\flo{straightforward parallelization (or static chunking, STATIC), self-scheduling (SS)~\cite{SS}, guided self-scheduling~\cite{GSS} (GSS), and factoring (FAC)~\cite{FAC}.}
In the present work, the scheduling behavior of the first {\em two} computational kernels using the above {\em four} scheduling techniques is reproduced 
to confirm that the implementations of the DLS techniques both, in simulation and in native codes, adhere to their original goals and specifications~\cite{FAC}. 

\alia{Confirming the adherence of the DLS implementation to their original design goals minimizes the sources of uncertainty in their implementation and helps avoiding unnecessary influences on the performance of scientific applications. 
For instance, a DLS that has been implemented to intensively use shared memory locks will cause unnecessary scheduling overhead. 
Therefore, adversely influencing performance. 
Another example would be that SS in present implementations performs differently than in the past for the particular implementation of MM considered in the past and in this work, leading to uncertainty whether the present SS implementation adheres to the original one. 
It was found that SS is properly implemented in the present and that the source of the discrepancy is due to the fact that MM in the past was compute-bound while in the present it is memory-bound. 
The achieved trust in the implementation of DLS techniques for shared-memory systems has already been transferred to their implementation for distributed-memory systems~\cite{Mohammed:2018a}.}
\flo{Confirming the adherence of the STATIC, SS, GSS, and FAC implementations to their original design goals lays the foundation for confirming the implementation of further DLS techniques that aim to better balance the increase in load balancing with the increase in scheduling overhead, such as, weighted factoring~\cite{WF}, adaptive weighted factoring~\cite{AWF}, and adaptive factoring~\cite{AF}.} 



\flo{Reproducibility is a key aspect of the scientific method~\cite{acmterminology}.
The reproduction of scientific experiments contributes to the validation of those experiments and to establish that the conclusions drawn from these experiments are of scientific relevance~\cite{repscascha}. 
In the present work, {\em reproduction}~\cite{acmterminology} is defined as revisiting a certain scientific problem, namely, the performance of DLS techniques~\cite{FAC}, without the original artifacts or the possibility to execute the artifacts on the original computing system~\cite{RP3}. 	
\alia{Reproduction is employed in this work as a means to attain and increase the trust in native and simulative implementations of DLS.}

The scheduling experiments selected from the work of Flynn Hummel {\em et al.}~\cite{FAC} \ali{were reproduced earlier~\cite{HPCC_FAC}, using simulative as well as native execution with DLS techniques implemented employing a centralized process coordination approach.
This work extends the reproduction of the experiments selected from the work of Flynn Hummel {\em et al.}~\cite{FAC} by investigating the implementation of the DLS techniques using decentralized process coordination. In the authors best understanding, DLS techniques were implemented originally using decentralized process coordination. For completeness both implementations, centralized and decentralized process coordination, are compared in this work.}
The simulative experiments were conducted with a simulator developed based on the SimGrid-SimDag (hereafter, \simdag{}) interface that employs individual representations of the two HPC platforms considered: IBM RP3 (past) and Intel Knights Landing (present).}
\flo{The reproduction of the selected scheduling experiments is a means for the {\em experimental verification} of the implementation of STATIC, SS, GSS, and FAC using~\simdag{}.}
Moreover, the selected scheduling experiments were performed natively on \ali{an Intel Knights Landing (hereafter, the KNL) architecture}, whose characteristics are captured in a platform file representation required by the simulator.
\simdag{} is then used to {\em predict} the performance of the execution on the KNL. 
The results of the native and simulative executions were compared and found in close agreement, which increases the confidence in the \mbox{simulation-based} prediction of the performance of DLS experiments.

The present work makes the following contributions:
(1)~Employs {\em reproduction} as a means to experimentally verify the \simdag{} implementation of STATIC, SS, GSS, and FAC by comparing the present simulation results with the corresponding results on the RP3 from the work of Flynn Hummel \mbox{{\em et al.}}~\cite{FAC}.
(2)~{\em Repeats} the selected scheduling experiments~\cite{FAC} on the KNL~7210 processor to explore whether conclusions of the past experiments hold on a modern computing system.
(3)~Introduces a \mbox{\simdag{}-based} simulator to simulate and predict the behavior of two computational kernels using four loop scheduling techniques~\cite{FAC} that employ the {\em decentralized} process coordination approach.
\flo{Experimentally verified implementations of DLS techniques can be useful for studying the relation between their use at different levels of scheduling~\cite{mls}. 
Moreover, the present work enables future studies on the scheduling behavior under various scheduling scenarios and in the presence of variable application and system properties.}

The remainder of this work is structured as follows:
\flo{The background on DLS techniques, the simulation toolkit, as well as an overview of relevant reproducibility studies are reviewed in  Section~\ref{sec:background}. 
The proposed methodology for performance reproduction and prediction of DLS is described in Section~\ref{sec:approach}. 
\ali{The reproduction of the selected experiments on the RP3 is presented in Section~\ref{sec:verification}.}
The reproduction of the selected scheduling experiments on the KNL architecture is detailed in Section~\ref{sec:repDLSKNL}.
The performance of the KNL-based experiments predicted with \simdag{} is compared against the performance of the native experiments on the KNL and discussed in Section~\ref{sec:repKNLsim}.
The conclusion and insights into future work are outlined in Section~\ref{sec:conc}.}

%% file: 2.tex
\section{Background and Related Work} 
\label{sec:background}
This section reviews the dynamic loop scheduling techniques and the \mbox{SimGrid} simulation toolkit. A number of relevant reproducibility studies are also discussed. 

\textbf{Dynamic loop scheduling.}
The loop scheduling techniques considered in this work can be classified into static and dynamic. 
\ali{Using straightforward parallelization (denoted STATIC)}, the parallel loop iterations are divided into equally-sized chunks. 
A processor is assigned {\em exactly one chunk} of iterations equal to the overall number of loop iterations~($N$) divided by the number of available processing elements~($P$). 
STATIC has a very low scheduling overhead~($h$), bounded above by~$P$.
Application performance may be degraded due to load imbalance if the execution of the loop iterations is characterized by high variability. 
\flo{\mbox{Self-scheduling}}~\cite{SS}~(SS), \ali{is a dynamic loop scheduling technique, at the other scheduling extreme,} whereby a processing element obtains a chunk consisting of {\em exactly one loop iteration} whenever it becomes available and requests work. 
When all loop iterations have been self-scheduled, the processors finish their execution at virtually the same time due to the fine-grain self-balancing of the workload. 
Scheduling a single loop iteration at a time leads to increased scheduling overhead over STATIC and potentially to an overall completion time greater than the optimal time. 

A number of other DLS techniques provide a \mbox{trade-off} between minimizing scheduling overhead and maximizing the load balancing. 
Two such techniques are guided \mbox{self-scheduling}~(GSS)~\cite{GSS} and factoring~(FAC)~\cite{FAC}. 
GSS assigns a chunk of loop iterations to an available and requesting processor that is equal to the number of the remaining unscheduled loop iterations~($R$) divided by the number of the processors~$P$.  
Therefore, chunks are of decreasing sizes, and workload can be balanced \ali{among} the processors also in the case of uneven processor start times. 
Even though GSS offers a good compromise between load balancing and scheduling overhead, it assigns a very large chunk to the first available worker. 
The execution of this chunk can dominate the application performance leading to load imbalance. 
FAC~\cite{FAC} is designed to balance the execution of loop iterations with variable execution times. 
It assigns chunks of loop iterations to available and requesting workers in batches vs. single loop iterations at a time, therefore, reducing the scheduling overhead~$h$. 
The number of the loop iterations in a chunk depends on the remaining number of loop iterations~$R$ and on the coefficient of variation (c.o.v.) of the loop iterations execution times. 

\textbf{\alia{Loop scheduling in simulation}.}
\mbox{SimGrid}~\cite{simgrid} is a scientific simulation framework for the study of the behavior of large-scale distributed computing systems, such as, the Grid, the Cloud, and peer-to-peer (P2P) systems. 
It provides ready-to-use models and application programming interfaces (APIs) to simulate various distributed computing systems. 
\mbox{SimGrid}~(hereafter, SG) provides \ali{four} different APIs for different simulation purposes. 
The \mbox{MetaSimGrid}~(MSG) and \mbox{SimDag}~(SD) provide APIs for the simulation of computational problems expressed as parallel independent tasks or as parallel task graphs, respectively.
The SMPI interface provides the functionality for \flo{the simulation of programs} written using the message passing interface (MPI) and targets developers interested in the simulation and debugging of their parallel MPI codes.
\ali{The newly introduced S4U interface currently supports most of the functionality of the MSG interface} \flo{with the purpose of also incorporating the functionality of the SD interface over time.}
This work considers the \simdag{} interface.  

%% file: 3.tex

\textbf{Related work.}
%
\flo{DLS techniques have previously been implemented in the \msg{} interface with the purpose of studying their scalability~\cite{mahad} and robustness against load imbalance~\cite{nitin}. 
Moreover, a number of DLS techniques were also implemented in \msg{} to study their resilience in a heterogeneous computing system~\cite{dlsmsg}. 
\ali{Another} closely related study performed \mbox{simulation-based} reproduction (using \msg{}) to confirm and validate the implementation of several DLS techniques in simulation~\cite{Hoffeins:2017b}. }


Two approaches can be employed to implement process coordination in the DLS techniques natively or in simulation: 
(1)~Centralized process coordination, using \flo{a} master-worker execution model; and
(2)~Decentralized process coordination, wherein each ``worker thread'' calculates and executes a chunk of work whenever it becomes available. 
\flo{A first effort to reproduce a selection of experiments from~\cite{FAC} using simulation considered the DLS techniques implemented using the master-worker execution model~\cite{HPCC_FAC}.}

\flo{The present work extends and complements previous work~\cite{HPCC_FAC} by investigating the reproduction of a selection of experiments~\cite{FAC} with the DLS techniques employing a decentralized process coordination approach using only ``worker threads'' without a ``master thread''.}
\flo{In this approach, threads (or processes) are responsible for obtaining work on their own from the central work queue \ali{shared via memory}, eliminating the master-side contention that characterizes the centralized \ali{process coordination} model.
The goal of this work is the use of performance~reproduction and performance~prediction as a means of {\em experimental verification} of the adherence of the DLS techniques implementation in \simdag{} to the original design goals and specifications. }


%% file: 4.tex
\section{Reproduction and Prediction Methodology}
\label{sec:approach}

In this work, four scheduling techniques: STATIC, SS, GSS, and FAC are implemented in \simdag{}. 
To confirm the implementation of these scheduling techniques, 
selected scheduling experiments from the original publication~\cite{FAC} are reproduced using simulation.
As mentioned earlier in Section~\ref{sec:background}, the DLS techniques under study can be implemented in one of two approaches:~(1)~Centralized process coordination;~(2) Decentralized process coordination. 
In this work, the DLS techniques implemented using the decentralized process coordination approach are investigated. In a centralized process coordination approach, the master calculates and assigns chunks of iterations to available and requesting workers. Also, the master can be dedicated or act as a worker as well when there are no requests to serve from workers. The decentralized approach studied in the present work is close to the implementation described in \ali{the original work}~\cite{FAC}, where each thread calculates and obtains a chunk of work when it becomes available. Atomic operations are used instead of locks to optimize the implementation as proposed in the original publication. The results of simulating and executing the scheduling experiments using centralized (from previous work~\cite{HPCC_FAC}) and decentralized (from the present work) process coordination approaches are compared in Sections~\ref{sec:verification} and~\ref{sec:repDLSKNL}. 

\begin{figure}[b]
\centering
\includegraphics[clip, trim=1cm 1cm 1cm 1cm,scale=0.75]{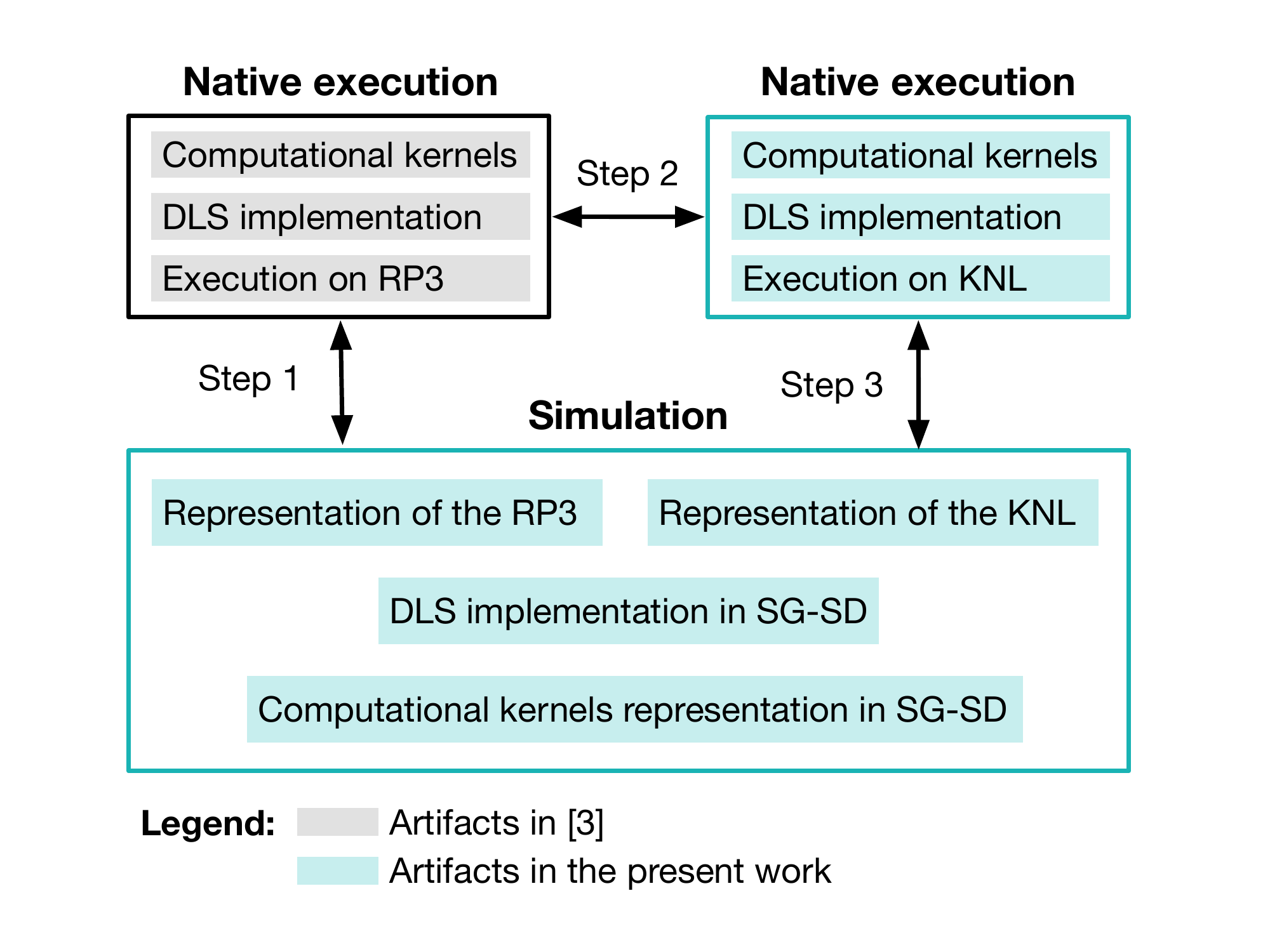}
\caption{\ali{Proposed reproduction and prediction methodology.}}
\label{figvvapproach}
\end{figure}

The reproduction \ali{and prediction approach consists of three steps} as illustrated in \figurename{~\ref{figvvapproach}}. 
Step~1 of the {\em reproduction} process is described in Section~\ref{sec:verification}. 
The results of the simulated experiments are compared with the results from the original publication~\cite{FAC} to confirm the present implementation of the DLS techniques in the \simdag{} simulator. The results of the original paper~\cite{FAC} were extracted from the figures using web plot digitizer\footnote{https://apps.automeris.io/wpd/}.\ali{Comparing the reproduced results with the original results ensures that the DLS techniques are delivering the same performance as in the original publication, and hence the verification of their implementation. The poor implementation of the DLS techniques may lead to load imbalances that should have been avoided using the DLS techniques.}

The selected DLS experiments are {\em reproduced} on the state-of-the-art manycore processor architecture, \mbox{the KNL}, as shown in Step~2 in \figurename{~\ref{figvvapproach}}.
The \mbox{KNL} is representative of modern manycore architectures that exhibit a high degree of parallelism, rendering it, therefore, an interesting architecture for the study of DLS. 
The details of the implementation and the reproduction of the DLS experiments on the KNL are presented in Section~\ref{sec:repDLSKNL}. 
This allows to examine whether the conclusions drawn from the DLS experiments described in \ali{the original work}~\cite{FAC} are influenced by the underlying system.
One can also examine whether the advancements in computer systems over three decades alter the conclusions of publications of the past.

The \simdag{} simulator is \ali{configured} (cf. Section~\ref{sec:repKNLsim}) to {\em predict} the performance of the selected experiments on the KNL architecture instead of the RP3 system denoted by Step~3 in \figurename{~\ref{figvvapproach}}.
The native execution results from Section~\ref{sec:repDLSKNL} are compared with the results of the simulated execution from Section~\ref{sec:repKNLsim}  to attain trustworthiness in the \mbox{\simdag{}-based} prediction of the performance of the selected DLS experiments. 
\ali{The proposed reproduction methodology in \figurename{~\ref{figvvapproach}} can be used in other scheduling studies to confirm the implementation of scheduling techniques (Step 1) and confirm the correctness of the simulation (Step 2).}

\paragraph*{Selection of the DLS Experiments}
The original paper~\cite{FAC} compared the performance of executing three different computational kernels: matrix multiplication, adjoint convolution, and  Gauss-Jordan method on the RP3 system using four different scheduling techniques STATIC, SS, GSS, and FAC. 
Two matrix sizes were used as input for each of the three kernels. 
Two variations of the adjoint convolution kernel were considered: with increasing task sizes and with decreasing task size.
All scheduling experiments in~\cite{FAC} were performed on the RP3\footnote{Each processor had its own local memory, configured in a shared address space. Every processor-memory element was connected to other elements using the Omega network~\cite{RP3}.} system~\cite{RP3}. 

\begin{algorithm}[h]
	\caption{Parallel matrix multiplication (MM)}
	\label{algo:matmul}
	\LinesNumbered
	\SetKwInput{Input}{Input}
	\SetKwInput{Output}{Output}
	\Input{Matrices $A$ and  $B$ each of size $n\times n$}
	\Output{Matrix $C$ of size $n\times n$}
	\KwData{$A$, $B$, $n$}
	\KwResult{$C \leftarrow A\times B$ }
	\DontPrintSemicolon
	\SetAlgoLined
	
	\For(in parallel ){$k = 1$ : $n \times n$} {
		$ i \leftarrow k/n$\;
		$ j \leftarrow k - n \times (k - 1) / n$\;
		$C[i,j] \leftarrow 0$\;
		\For{$l = 1$ : $n$} {
			$C[i,j] \leftarrow C[i,j] + A[i,l] \times B[l,j]$\;
		}
	}
\end{algorithm}

\begin{algorithm}[h]
	\caption{Parallel adjoint convolution (AC-d)}
	\label{algo:adjointconv}
	\LinesNumbered
	\SetKwInput{Input}{Input}
	\SetKwInput{Output}{Output}
	\Input{Two matrices $A$ and  $B$ each of size $n\times n$}
	\Output{Matrix $C$ of size $n\times n$, where $C$ the adjoint convolution of $A$ and $ B$}
	\KwData{$A$, $B$, $C$, $n$, $const$}
	\DontPrintSemicolon
	
	\SetAlgoLined
	\For(in parallel){$k = 1$ : $n \times n $} {
		$C[k] \leftarrow 0$\;
		\For{$l = k$ : $ n \times n $} {
			$C[k] \leftarrow C[k] + const \times A[l] \times B[l-k]$\;
		}
	}
\end{algorithm}

The matrix multiplication (MM) and adjoint convolution with decreasing task sizes (AC-d) kernels are selected for reproduction and prediction in this work, with matrix sizes of $300\times300$ and $75\times75$, respectively.
The computational kernels are described in Algorithms~\ref{algo:matmul} and~\ref{algo:adjointconv}. 
Larger matrices are used in the scheduling experiments on the KNL to arrive at an execution cost on the KNL close to that of the experiments on the RP3 system. Using large matrices results in a longer program execution time. Errors in the execution time measurements and in the overhead are small compared to the measurement of the program execution. These timings are negligible, i.e., the time measurement function calls require 16.15 microseconds for a program execution time of 329~seconds. The matrix sizes of \mbox{$5500\times5500$} and \mbox{$600\times600$}  are used for the MM and the AC-d kernels on the KNL, respectively. 
The selected experiments details are summarized in Table~\ref{Fac_expr}.
\begin{table}[h]
	\centering
	\caption{\ali{Selected scheduling experiments}}
	\label{Fac_expr}
	\resizebox{\textwidth}{!}{
		\begin{tabular}{l@{\hskip 0.05in}|l@{\hskip 0.05in}|l@{\hskip 0.05in}|l}
			\hline
			\textbf{\begin{tabular}[l]{@{}l@{}}Computational kernel\end{tabular}}         & \textbf{\begin{tabular}[l]{@{}l@{}}Matrix \\ size
			\end{tabular}} & \textbf{\begin{tabular}[l]{@{}l@{}}Scheduling\\ method\end{tabular}}              & \textbf{\begin{tabular}[l]{@{}l@{}}Number of \\ processors\end{tabular}}               \\ \hline
			Matrix multiplication (MM)                 & $300\times300$                                                        & \multirow{2}{*}{\begin{tabular}[]{@{}l@{}}STATIC, SS\\ GSS, FAC\end{tabular}} & \multirow{2}{*}{\begin{tabular}[l]{@{}l@{}}4, 8, 16, 24,\\ 32, 40, 48, 56\end{tabular}} \\ \cline{1-2}
			\begin{tabular}[l]{@{}l@{}}Adjoint convolution \ali{with}\\ decreasing task sizes (AC-d)\end{tabular}  & $75\times75$                                                          &                                                                                   &                                                                                        \\ \hline
		\end{tabular}
	}
\end{table}

The two kernels represent two different task granularities: equal task sizes and decreasing task sizes. 
Each iteration of a kernel's \emph{for~loop} was considered a task to be scheduled. 

 
The open-source simulator and the raw results obtained from simulated and native executions for this work are available online~\cite{HPCS_repo}.
An Easybuild\footnote{http://easybuild.readthedocs.io} configuration file is also provided to ensure the creation of an experimental environment that is similar to the one used for this work.




%% file: 5.tex
\section{Reproduction of Selected Experiments via Simulation} 
\label{sec:verification}

To confirm the implementation of the four scheduling techniques in \simdag{}, the selected scheduling experiments on the RP3 system~\cite{FAC} are reproduced and compared with those obtained using \simdag{}~\cite{simgrid}. 
Every iteration of a kernel's outer loop is modeled as a \simdag{} \emph{sequential computation task}. 
The amount of work contained in each computational task is specified in number of floating point operations (FLOP) in the simulator.
For both MM and AC-d, the FLOP count in each iteration is inferred from their pseudocodes.
This number is used in the simulator as the amount of work in each sequential computation task. The DLS techniques are implemented using decentralized \ali{process coordination}. The pseudocode of the \simdag{} simulator of the parallel execution of the two kernels with DLS techniques is listed in Algorithm~\ref{algo:simgrid1}. 
%
\begin{algorithm}[H]
	\caption{\simdag{} pseudocode \textemdash{} decentralized process coordination}
	\label{algo:simgrid1}
	\DontPrintSemicolon
	\LinesNumbered
	\SetAlgoLined
	\SetKwInput{Input}{Input}
	\SetKwInput{Output}{Output}
	\Input{$platformFile$, $numThreads$, $kType$, $pSize$, $method$}
	\Output{simulatedTime}
	\KwData{$schedulingStep$, $scheduledTasks$, $chunkSize$, $hosts$, $tasks$, $numTasks$, $dummyTask$, $dummyComm$, $changedTasks$}
	$numTasks \leftarrow pSize \times pSize$\;
	$tasks \leftarrow CreateTasks(kType,pSize)$\;
	\ForEach{$i \in numTasks$}
	{
		SD\_task\_watch($tasks[i]$, SD\_DONE)\;
	}
	\tcc{Create a computational task to represent create threads overhead}
	$dummyTask \leftarrow $ SD\_task\_create\_comp\_seq(``createThreads'', thread creation overhead according to $numThreads$ in FLOPs)\;
	SD\_task\_schedule($dummyTask$, $hosts[0]$)\; 

\addtocounter{algocf}{-1}
\end{algorithm}

\clearpage

\begin{algorithm}[H]
	\caption{\simdag{} pseudocode \textemdash{} decentralized process coordination - \mbox{continued}}
	\label{algo:simgrid1_2}
	\DontPrintSemicolon
	\LinesNumbered
	\SetAlgoLined
	\setcounter{AlgoLine}{7}
	\tcc{Run the simulation until a task is completed}
\While{!(is\_empty($changedTasks$=SD\_simulate(-1)))}{
	\For{$i = 0$ : $numThreads$}{
		\If{($scheduledTasks < numTasks$) and is\_free($hosts[i]$)}{
			$chunkSize   \leftarrow$  calculate\_chunk\_size($numTasks$, $numThreads$, $schedulingStep$, $method$)\;
			\tcc{Create a scheduling overhead task according to the scheduling method}
			$dummyTask \leftarrow $ SD\_task\_create\_comp\_seq(``scheduling overhead'', scheduling overhead in FLOPs correponding to $method$)\;
			
			$dummyComm \leftarrow $ SD\_task\_create\_end\_end\_comm(``assigning chunk'', $chunkSize \times pSize \times 8$)\;
			\tcc{Add dependencies between calculating chunk overhead, assigning overhead and the start of the execution of the chunk of tasks}
			SD\_task\_schedule($dummyTask$, $hosts[i]$)\;
			Schedule\_comm\_A\_to\_B($dummyTask$, $hosts[0]$, $hosts[i]$)\;
			\tcc{Schedule the chunk of tasks}
			\For{$j = 0$ : $chunkSize$}{
				SD\_task\_schedule($tasks[scheduledTasks]$, $hosts[i]$)\;
				Increment $scheduledTasks$\;
				
			}
			Increment $schedulingStep$\;
		}
	}
}
Print the simulated time\;
Terminate the program\;
\end{algorithm}

A \simdag{} \emph{sequential computation} task is created to represent the scheduling overhead of each DLS technique. This task is scheduled on the available thread in each simulated scheduling round. 
The amount of work performed by each of these scheduling tasks varies and depends on the selected scheduling technique. 
The values for the amount of work performed by each of these scheduling overhead tasks are obtained empirically, to match the simulation results to the results in the original publication~\cite{FAC}. Specifically, they are found to be 75, 400, 750, and 750 FLOP for STATIC, SS, GSS, and FAC techniques, respectively. 
A \simdag{} \emph{end-to-end communication task} is also created in each scheduling round to simulate the time taken to send the assigned chunk of tasks from process~0, to the available process that needs work. It is assumed that, initially, process~0 stores all the data, and other processes transfer one column of the matrix from process~0 for every task they obtain. This data strategy is referred to as \emph{pool of tasks and data}.

The amounts for computation (FLOP) and communication~(Byte) in each loop iteration for the two selected computational kernels are presented in Table~\ref{appparams}. 
Two factors, $g_1$ and $g_2$, are used to capture the unknown effects in the execution of the computational kernels on the RP3 system.
These factors cover all software- and hardware-related aspects that may influence program execution on RP3, e.g., memory system and operating system interference. These factors are presented in Table~\ref{appparams}, are unitless and are experimentally determined to be 35 and 60, respectively. 
\begin{table}[b]
	\centering
	\caption{Computational kernels parameters for their simulation on the RP3 system.}
	\label{appparams}
	\resizebox{\textwidth}{!}{
		\begin{tabular}{l@{\hspace{0.2cm}}l@{\hspace{0.2cm}}l@{}}
			\hline
			\textbf{\begin{tabular}[l]{@{}l@{}}Computational \\kernel\end{tabular}}  &  	\textbf{\begin{tabular}[l]{@{}l@{}}Task size (FLOP)\end{tabular}}  &  \textbf{\begin{tabular}[l]{@{}l@{}}Communication size (Byte)\end{tabular}} \\ \hline
			\begin{tabular}[l]{@{}l@{}} MM \end{tabular} &  \begin{tabular}[l]{@{}l@{}}$g_1 \times  (5+2 \times \text{rowLength})$\end{tabular}    &  \begin{tabular}[l]{@{}l@{}}$ \text{chunkSize} \times \text{rowLength}$\end{tabular}    \\ 
			\begin{tabular}[l]{@{}l@{}}AC-d \end{tabular} 	& 	\begin{tabular}[l]{@{}l@{}}$g_2 \times 3 \times (\text{matrixSize}-\text{iterationID})$\end{tabular}     & 	\begin{tabular}[l]{@{}l@{}}$ \text{chunkSize} \times \text{rowLength}$\end{tabular} \\  \hline
		\end{tabular}
	}
\end{table}

%

To provide the SimGrid simulation engine with the specifications of the simulated system, it requires an XML file as a \textit{platform file}. 
Each processor in the RP3 system is represented as a host in the SimGrid \textit{platform file} used in the reproduction experiments. All hosts (processors) are interconnected by creating a communication link between every host and all others. 
Additional details about the RP3 system are extracted from \ali{the work that introduced the RP3 system}~\cite{RP3}, such as processor speed (1.562~MFLOP/s), network bandwidth (50~Mbit/s), and latency (2~$\mu$s).

All simulations are performed using \simdag{}~3.16 on a manycore compute node with an Intel~KNL~processor (7210) running at~1.3~GHz, using CentOS~operating~system, version~7.2.1511.
The GNU~C compiler, version~6.3.0, is used for the compilation of the simulator with~\flo{\texttt{-g}~\texttt{-Wall}} as compilation flags.
%

\begin{landscape}
\begin{figure}[t]
	\begin{adjustbox}{minipage=\linewidth,frame}
	\centering
	\subfloat[STATIC \textemdash{} MM]{%
		\includegraphics[scale=0.3, clip,trim=0cm 10cm 18cm 0cm]{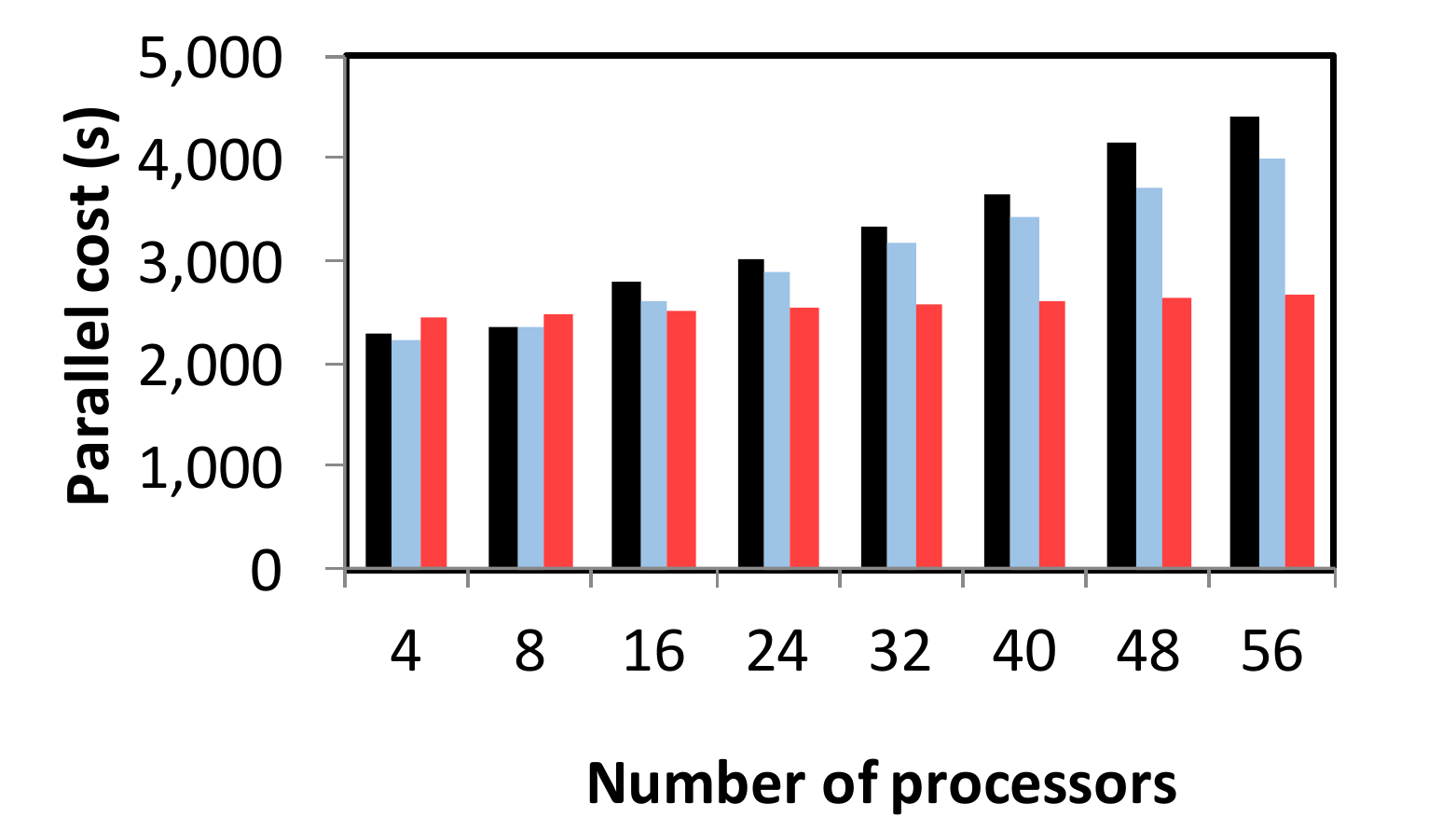}%
		\label{fig2:left}%
	} 
	\subfloat[SS \textemdash{} MM]{%
		\includegraphics[scale=0.3, clip,trim=0cm 10cm 18cm 0cm]{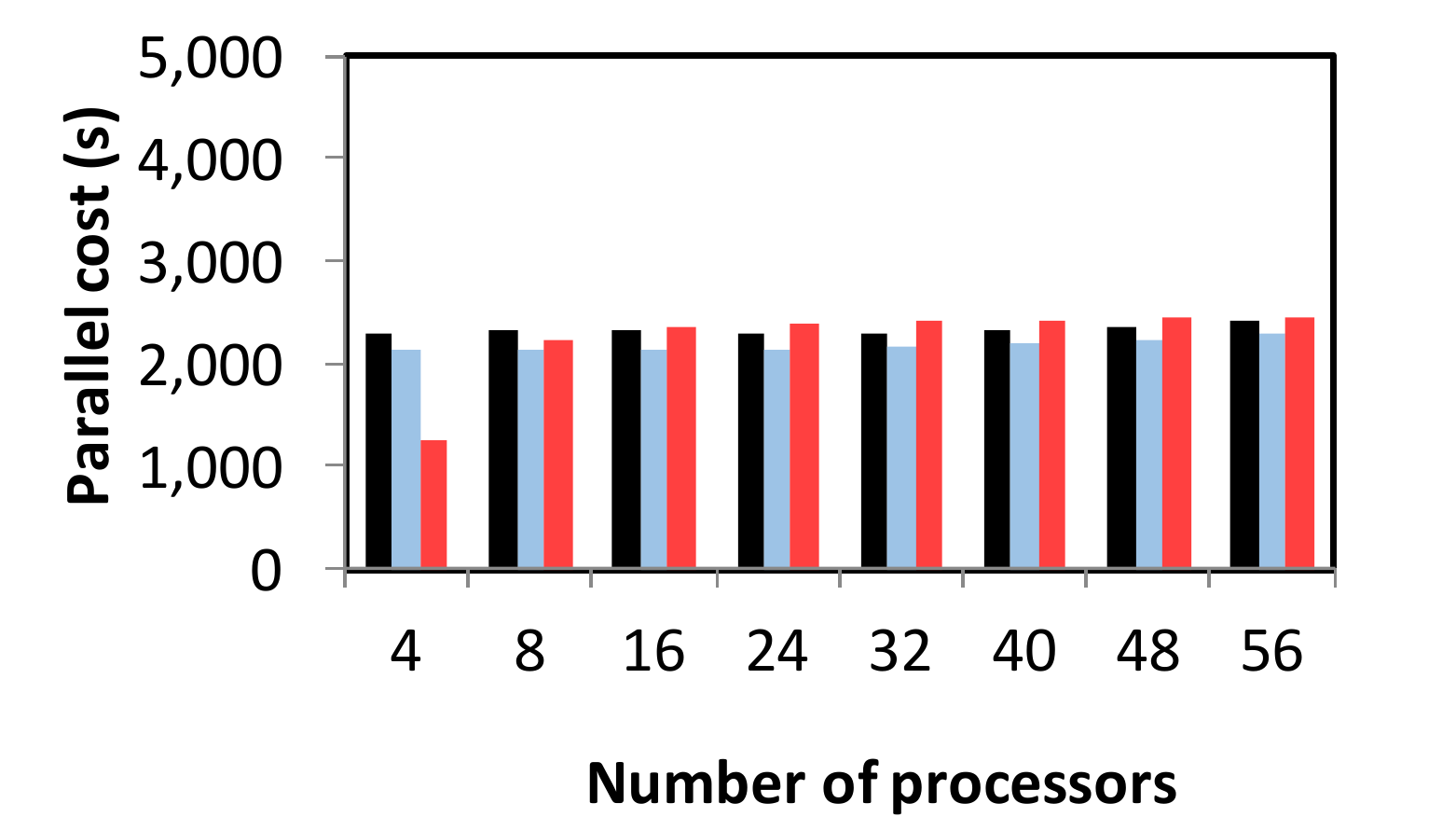}%
		\label{fig2:left2}%
	} 
	\subfloat[GSS \textemdash{} MM]{%
		\includegraphics[scale=0.3, clip,trim=0cm 10cm 18cm 0cm]{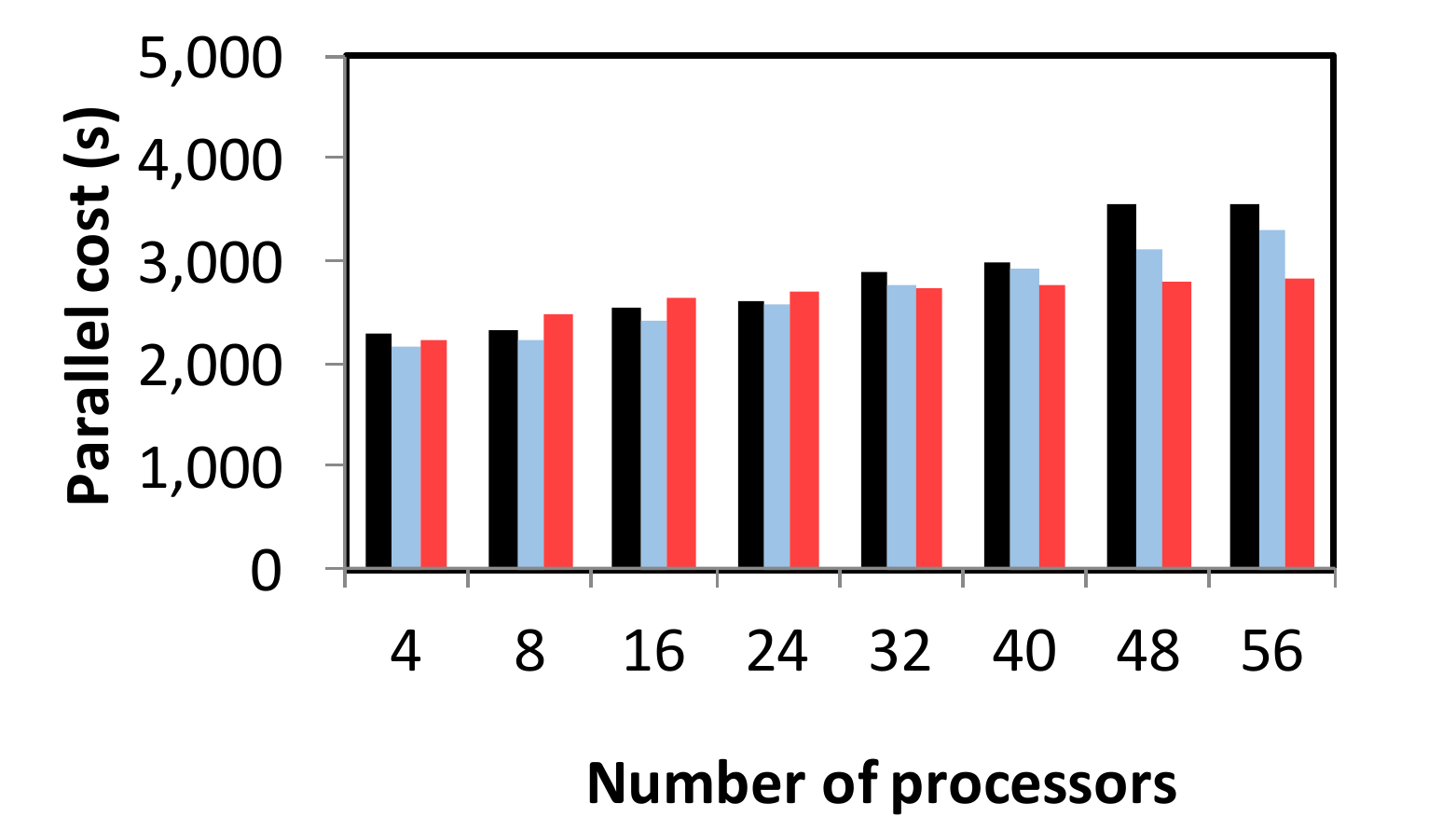}%
		\label{fig2:left3}%
	} 
	\subfloat[FAC \textemdash{} MM]{%
		\includegraphics[scale=0.3, clip,trim=0cm 10cm 18cm 0cm]{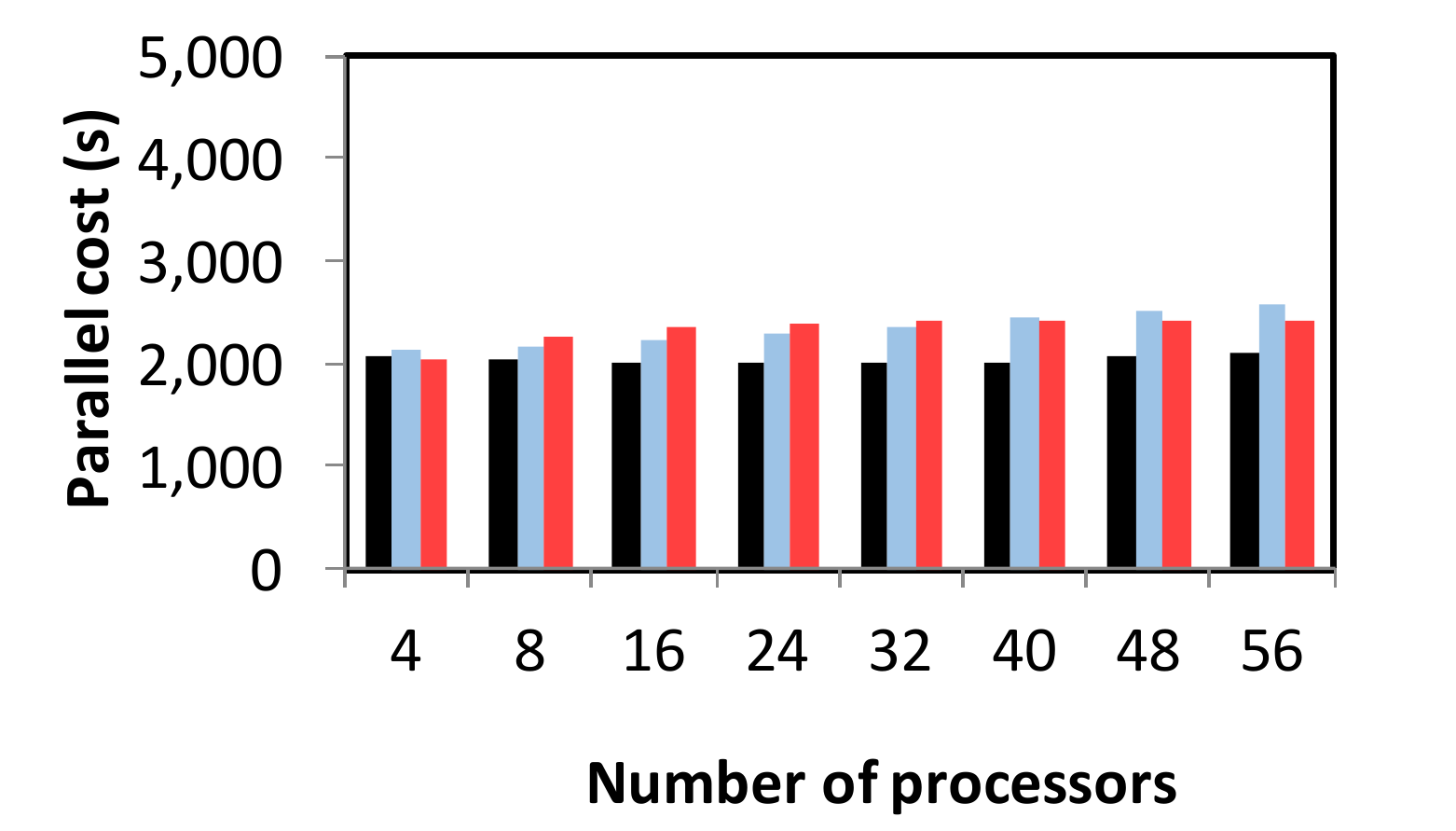}%
		\label{fig2:left4}%
	} 
	\\
	\subfloat[STATIC \textemdash{} AC-d]{%
		\includegraphics[scale=0.3, clip,trim=0cm 10cm 18cm 0cm]{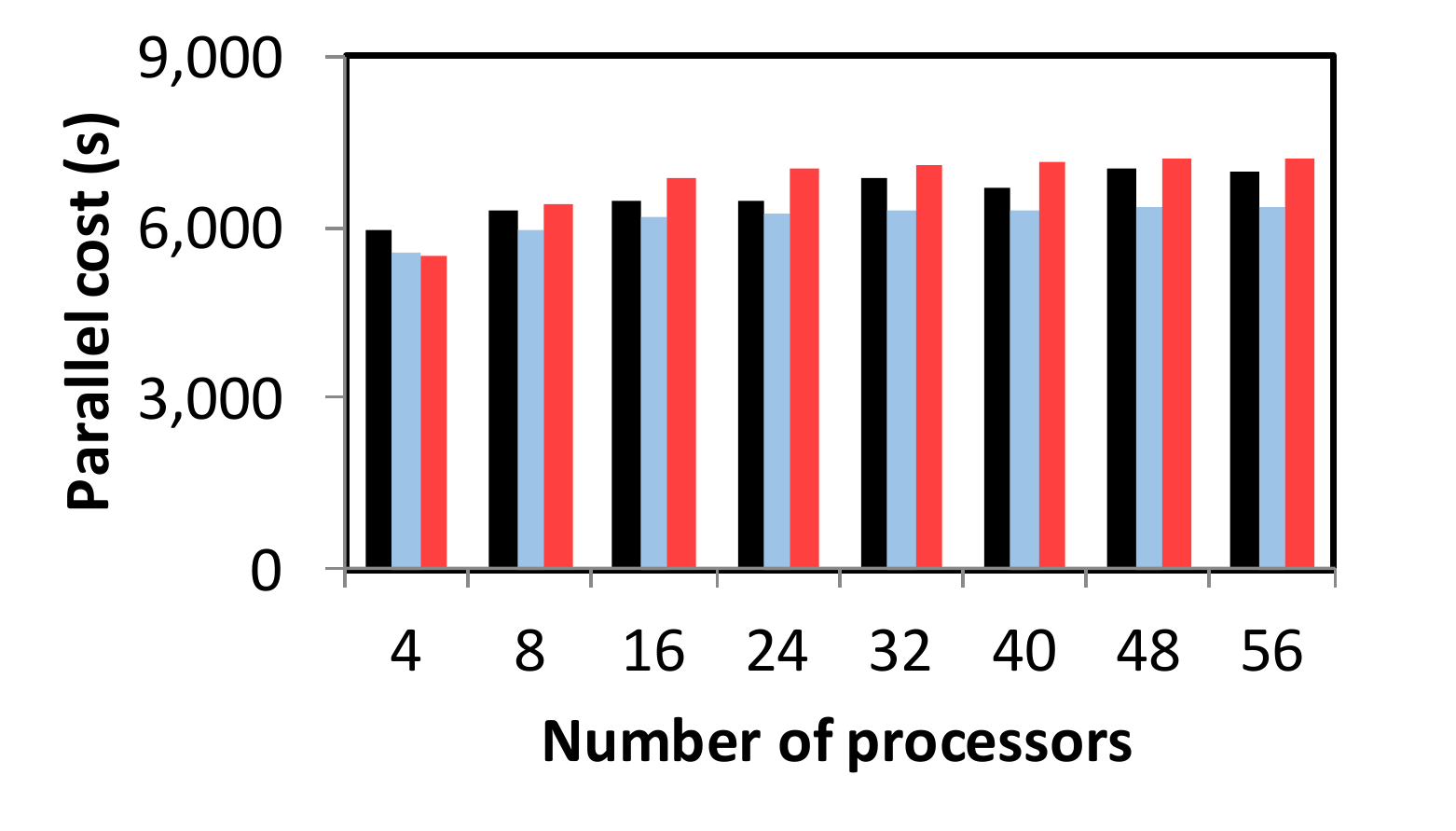}%
		\label{fig2:middle}%
	}
	\subfloat[SS \textemdash{} AC-d]{%
		\includegraphics[scale=0.3, clip,trim=0cm 10cm 18cm 0cm]{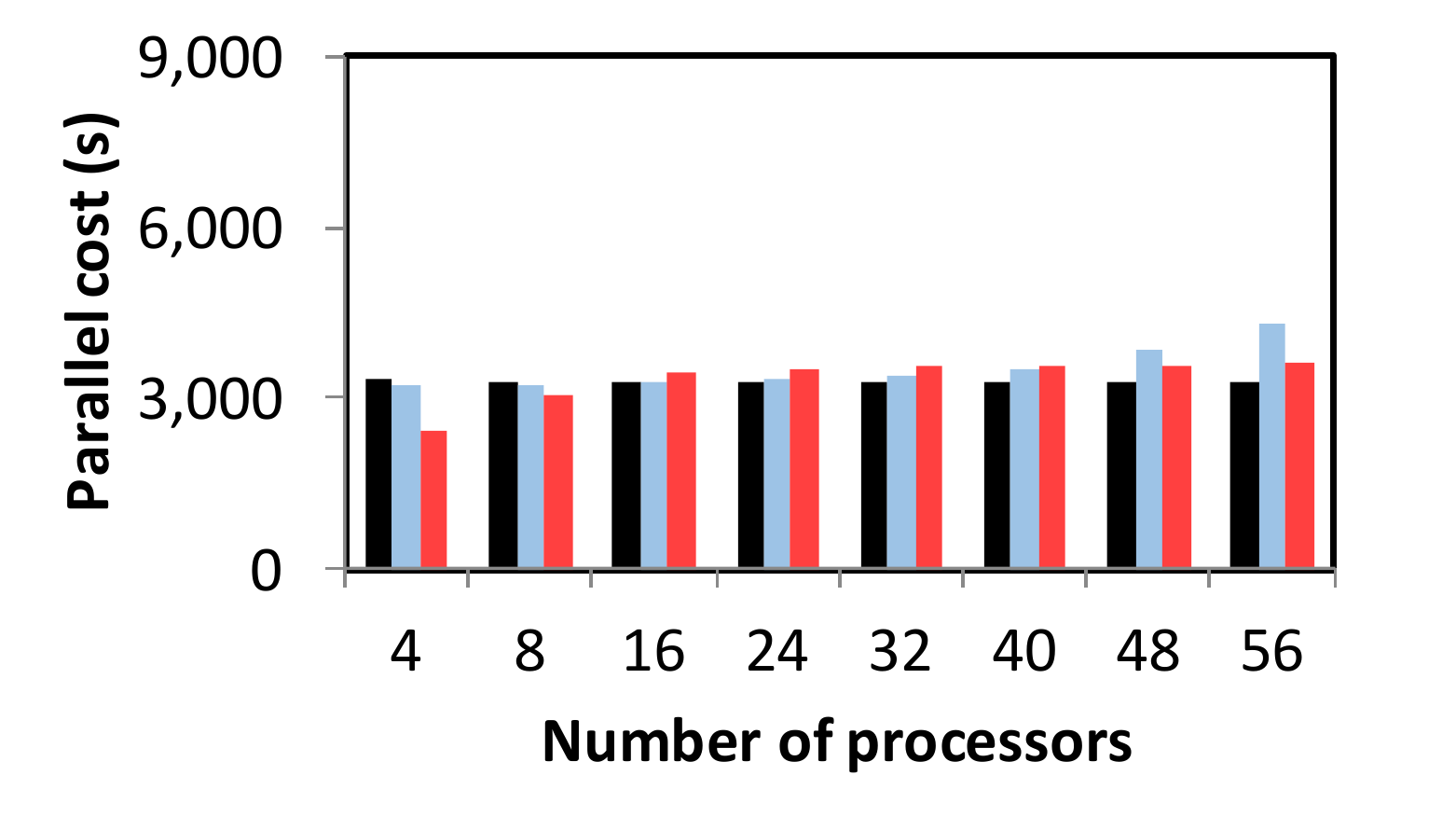}%
		\label{fig2:middle2}%
	}
	\subfloat[GSS \textemdash{} AC-d]{%
		\includegraphics[scale=0.3, clip,trim=0cm 10cm 18cm 0cm]{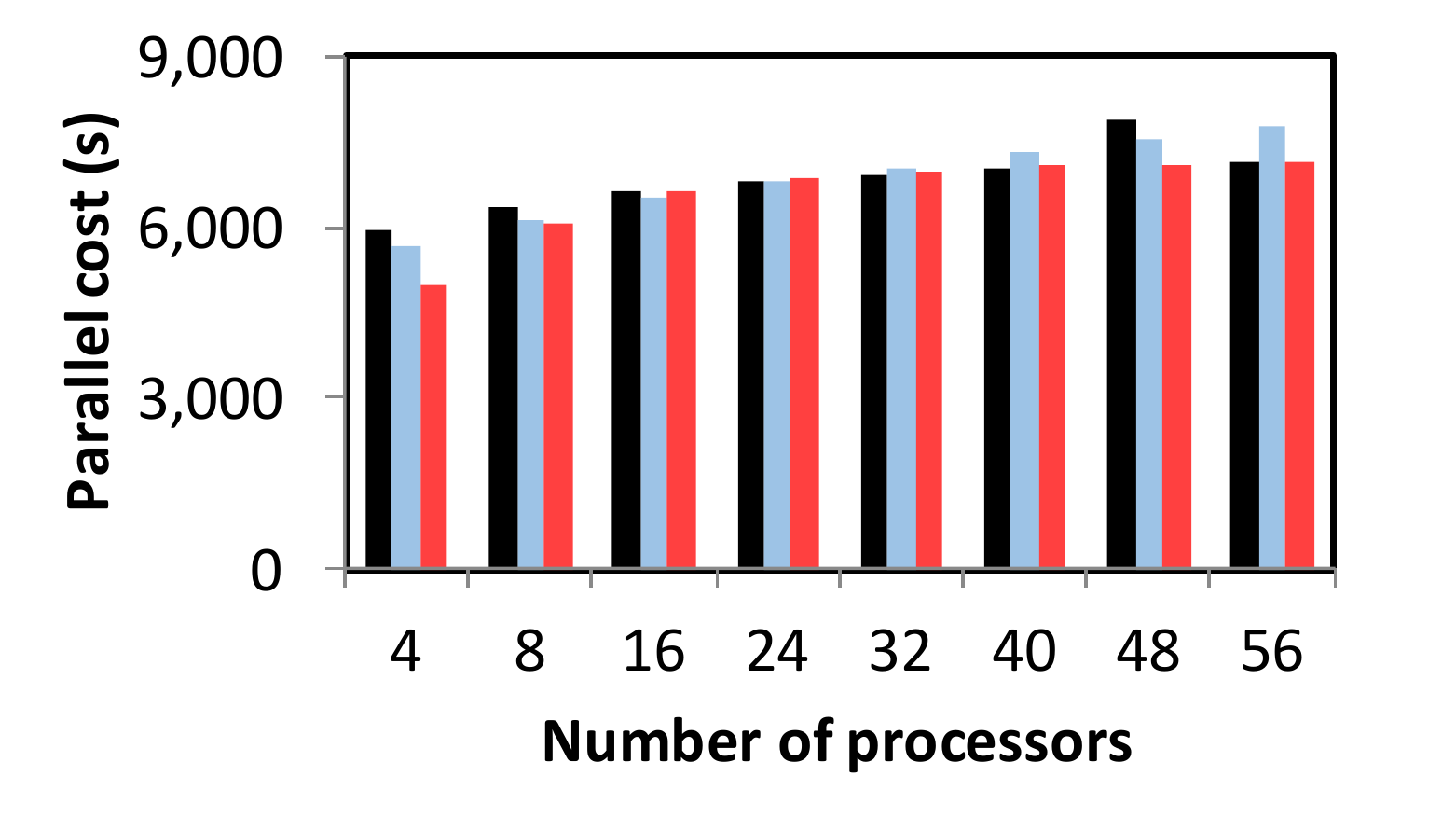}%
		\label{fig2:middle3}%
	}
	\subfloat[FAC \textemdash{} AC-d]{%
		\includegraphics[scale=0.3, clip,trim=0cm 10cm 18cm 0cm]{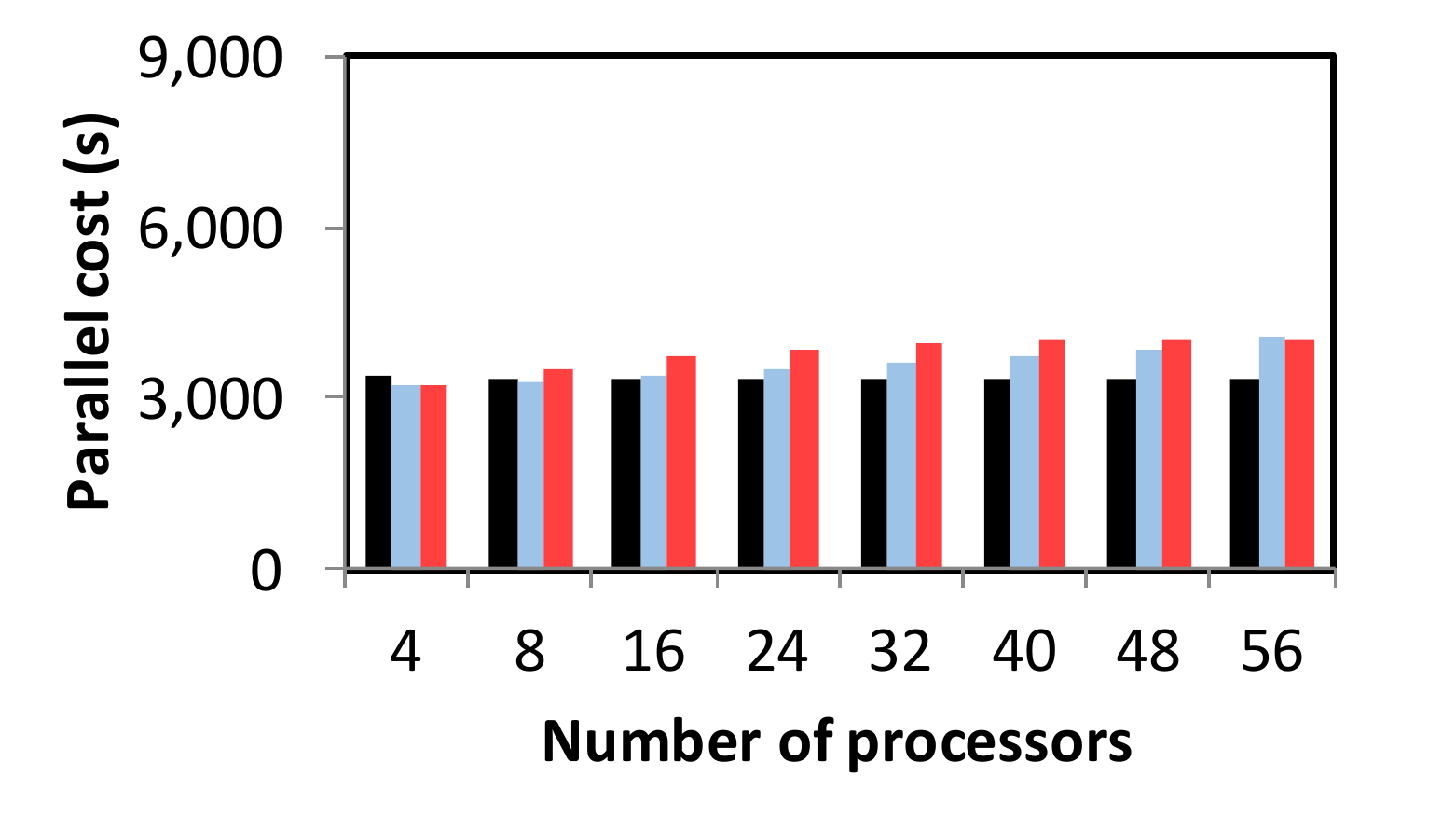}%
		\label{fig2:middle4}%
	}\\
\subfloat{%
	\includegraphics[scale=0.50, clip,trim=0cm 16.5cm 0cm 1cm]{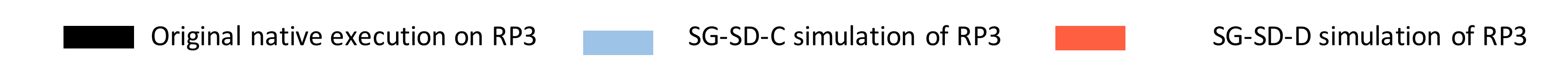}%
	\label{fig2:legend}%
}
\end{adjustbox}
	\caption{\alia{Simulation} results \alia{for} the selected DLS experiments on the RP3 system using a decentralized process coordination (\simdag{}-D) obtained with \simdag{} (red bars) compared with the simulation results \alia{for} the selected DLS experiments on the RP3 system using a centralized process coordination (\simdag{}-C)~\cite{HPCC_FAC} (blue bars) and the original publication~\cite{FAC} results (black bars). 
		\mbox{Parallel cost = parallel~program~execution~time~$\times$~number~of~threads.}} 
	\label{figRP3_diff}
\end{figure}   
\end{landscape}

%
\paragraph*{Results of Reproduction}

\alia{The selection of parallel cost as a performance metric (over the parallel execution time) is due to the fact that the parallel cost was used in the original publication~\cite{FAC} that this work compares against. The parallel cost reflects the sum of the time that each processing element spends solving the problem~\cite{kumar1994introduction}.}
The simulative performance for executing MM and AC-d using a decentralized \ali{coordination} approach with \simdag{} (\mbox{\flo{SG-SD-D}}) compared against the original native performance results~\cite{FAC} is illustrated in \figurename{~\ref{figRP3_diff}}. 
These results show that the simulation performance is close to the native performance in the original publication. 
The simulative performance of the same experiments using a centralized process coordination~\cite{HPCC_FAC} (\mbox{\flo{SG-SD-C}}) is also compared against the original native performance results~\cite{FAC} in \figurename{~\ref{figRP3_diff}}. 
\flo{The percent error ($\%E$) between the simulative execution time in this work ($T_{sim}$) and the original native execution time ($T^{o}_{nat}$)~\cite{FAC} is calculated as: }
$\%E= \left(1-\frac{T_{sim}}{T^{o}_{nat}}\right) \times 100$.


 A \alia{positive} \flo{percent error $\%E$} indicates that the simulator underestimates the original execution time, while a \alia{negative} $\%E$ signifies overestimation.
The minimum absolute $\%E$ between \simdag{}-D and the native execution is $0.073\%$, for \mbox{GSS~\textemdash{}~AC-d} and 56~threads, 
as can be observed from \figurename{~\ref{figRP3_diff}}(g).
The maximum absolute $\%E$ is $45.89\%$ in the case of \mbox{SS \textemdash{} MM} and 4 threads, 
as can be observed from \figurename{~\ref{figRP3_diff}}(b). 
The average of the absolute $\%E$ is $10.89\%$ in all the scheduling experiments on the RP3 system and the \simdag{} simulation results shown in the present work.
For the results \ali{centralized process coordination}~\cite{HPCC_FAC}, the minimum and the maximum absolute $\%E$ are 0.49\%, and 30.94\%, respectively, in the case of \mbox{GSS \textemdash{} AC-d} and 24 threads (see \figurename{~\ref{figRP3_diff}}(g)) and \mbox{SS \textemdash{} AC-d} and 56 threads (see \figurename{~\ref{figRP3_diff}}(f)).
The average of the absolute $\%E$  is $7.44\%$ between the simulative results~\cite{HPCC_FAC} (\simdag{}-C) and the native execution results~\cite{FAC}.
\flo{The simulative results follow a similar trend to the original native experiments, which is of high relevance for the comparison of different scheduling techniques.}
\textbf{These results \emph{confirm} that \alia{the} implementation of the considered DLS techniques in \simdag{} adheres to \alia{their} implementation used in the original publication~\cite{FAC}}.

%% file: 6.tex
\section{Reproduction of Selected Experiments via Native Execution}
\label{sec:repDLSKNL}

\ali{For the purpose of reproducing the selected experiments~\cite{FAC} via native execution in the present work, the two computational kernels were implemented in C.}
Their parallelization considers the scheduling techniques STATIC, SS, GSS, and FAC using Pthreads~\cite{pthreads}.
The Pthreads threading library is chosen due to its lightweight threading and its efficiency in communication and data exchange on shared memory computing systems.  

A decentralized process coordination is used in parallelizing the computational kernels with the DLS techniques. The main thread, thread 0, creates a number of threads equal to the number of cores in the current experiment minus one (the main thread). All threads, including the main thread, execute the code described in Algorithm~\ref{algo:native_code}. Each thread obtains work using the \texttt{obtain\_work} function described in Algorithm~\ref{algo:get_work}. The program holds two global variables that represent the current state of the program: \texttt{schedulingStep} and \texttt{currentIndex}. The \texttt{currentIndex} represents the loop index of the outer loop that is parallelized of the computational kernels and indicates the program progress. The \texttt{obtain\_work} function updates these two variables after each work assignment to advance the program state. Updates to these global variables (Lines 1 and 3 in Algorithm~\ref{algo:get_work}) are performed using \emph{atomic} operations to avoid data races between parallel threads. 
The size of the allocated chunk of work is calculated by the selected loop scheduling technique. 
All  threads are pinned on the cores of the experiments platform using the scatter strategy, to ensure better and more stable performance among execution runs. 
 

\begin{algorithm}[]
	\caption{Decentralized DLS \textemdash{} \mbox{thread execution}}
	\label{algo:native_code}
	\DontPrintSemicolon
	\LinesNumbered
	\SetAlgoLined
	\SetKwInput{Input}{Input}
	\SetKwInput{Output}{Output}
	\SetKwInput{Global}{Global data}
	\SetKwInput{Local}{Local data}
	\Input{$theadID$}
	\Output{$void$}
	\Global{$method$, $schedulingStep$, $currentIndex$}
	\Local{$start$, $chunkSize$}
	\While{True}
	{
		\uIf{obtain\_work($start$, $chunkSize$)}{
			execute\_kernel($start$, $chunkSize$)\;
		}
		\Else{
			break\;
		}
	}
	\tcc{Exit thread}
	return $NULL$\;
\end{algorithm}

\begin{algorithm}[]
	\caption{Decentralized DLS \textemdash{} \mbox{obtain work} }
	\label{algo:get_work}
	\DontPrintSemicolon
	\LinesNumbered
	\SetAlgoLined
	\SetKwInput{Input}{Input}
	\SetKwInput{Output}{Output}
	\SetKwInput{Global}{Global data}
	\SetKwInput{Local}{Local data}
	\Input{$method$, $schedulingStep$, $currentIndex$}
	\Output{$start$, $chunkSize$}
	\Global{$method$, $schedulingStep$, $currentIndex$, $numTasks$}
	\Local{$myStep$}
	
	$myStep \leftarrow$ fetch\_and\_add($schedulingStep$, 1)\;
	$chunkSize \leftarrow$ calculate\_chunk($numThreads$, $numTasks$, $myStep$)\;
	$start \leftarrow$  fetch\_and\_add($currentIndex$, $chunkSize$)\;
	\uIf{$start < numTasks$}{
		\uIf{$start+chunkSize >= numTasks$}{
			$chunkSize \leftarrow numTasks - start$
	}
		return True\;
	}
	\Else{
		return False\;
	}
	
\end{algorithm}

The parallel cost is reported for each experiment.
This cost is calculated as the product of the program's parallel execution time and the number of threads. 
A script to run the experiments and calculate the confidence interval is used to execute each experiment for a minimum number of 20 times and maximum of 100. For all the experiments, the script stops after 20 times yielding a confidence interval of less than 5\% and a confidence level of 95\%. 

\begin{landscape}
\begin{figure*}[]
		\begin{adjustbox}{minipage=\linewidth,frame}
		\centering
	\subfloat[STATIC \textemdash{} MM]{%
		\includegraphics[scale=0.3, clip,trim=0cm 10cm 18cm 0cm]{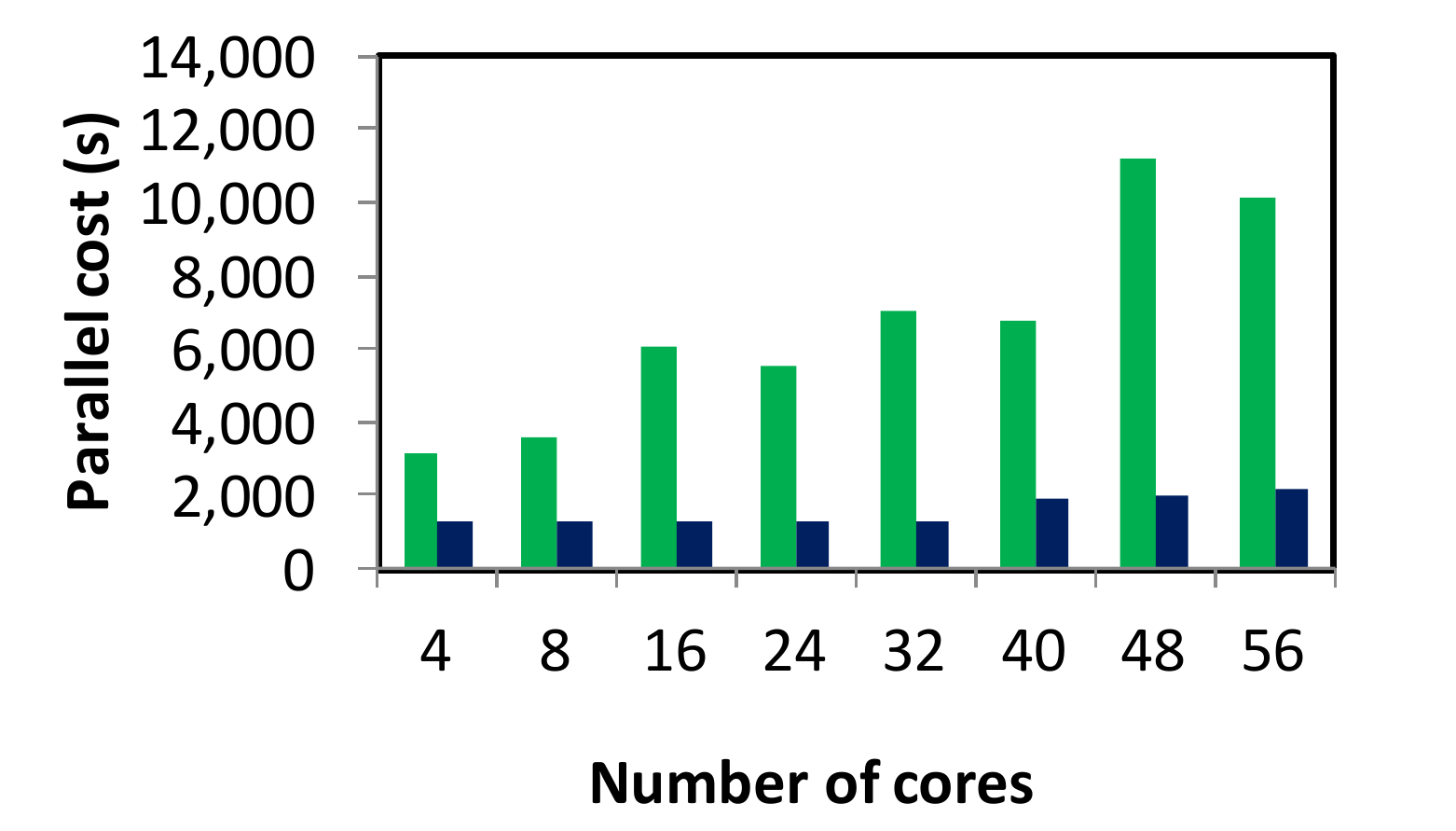}%
		\label{fig3:left}%
	} 
	\subfloat[SS \textemdash{} MM]{%
		\includegraphics[scale=0.3, clip,trim=0cm 10cm 18cm 0cm]{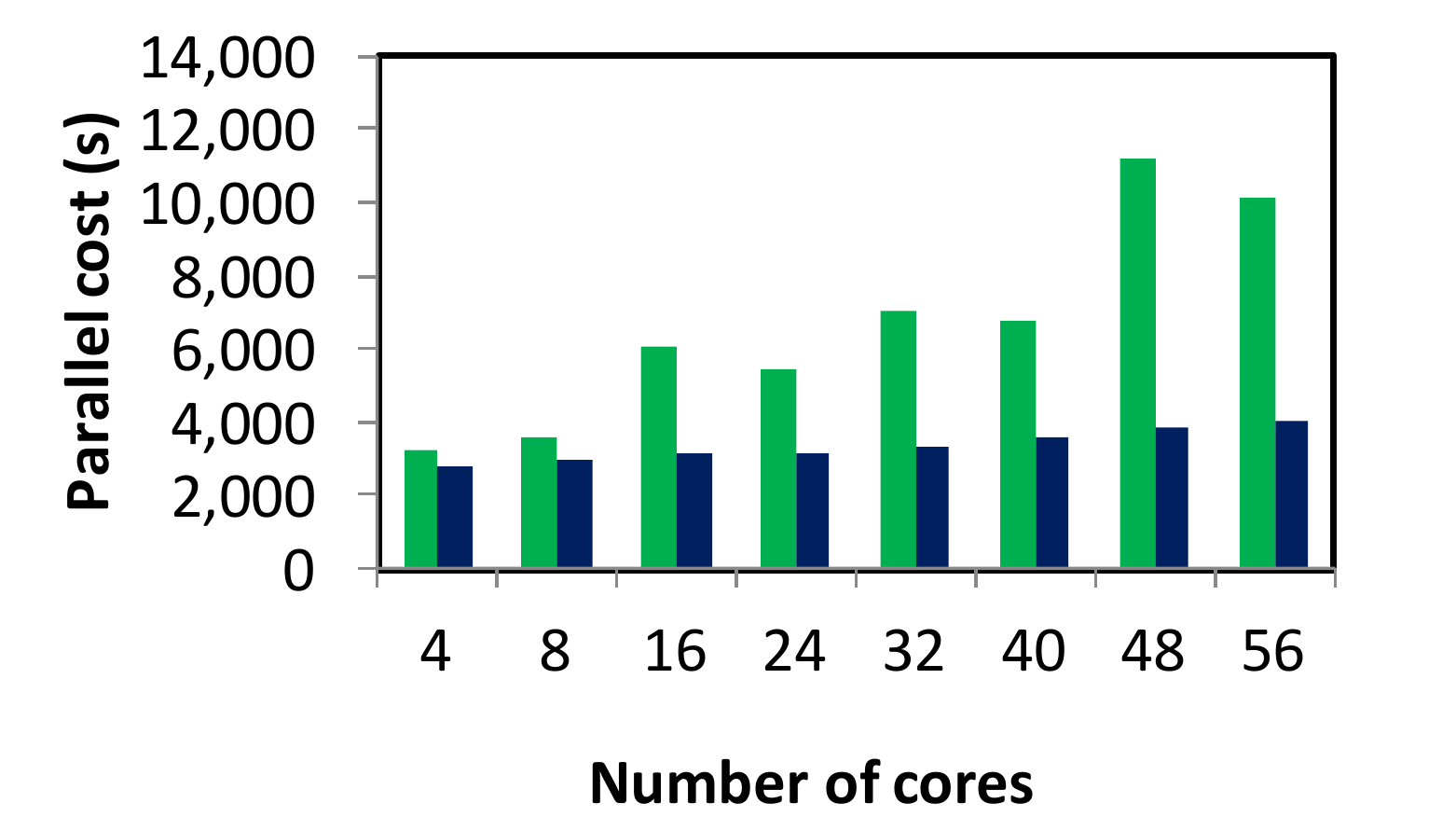}%
		\label{fig3:left2}%
	} 
	\subfloat[GSS \textemdash{} MM]{%
		\includegraphics[scale=0.3, clip,trim=0cm 10cm 18cm 0cm]{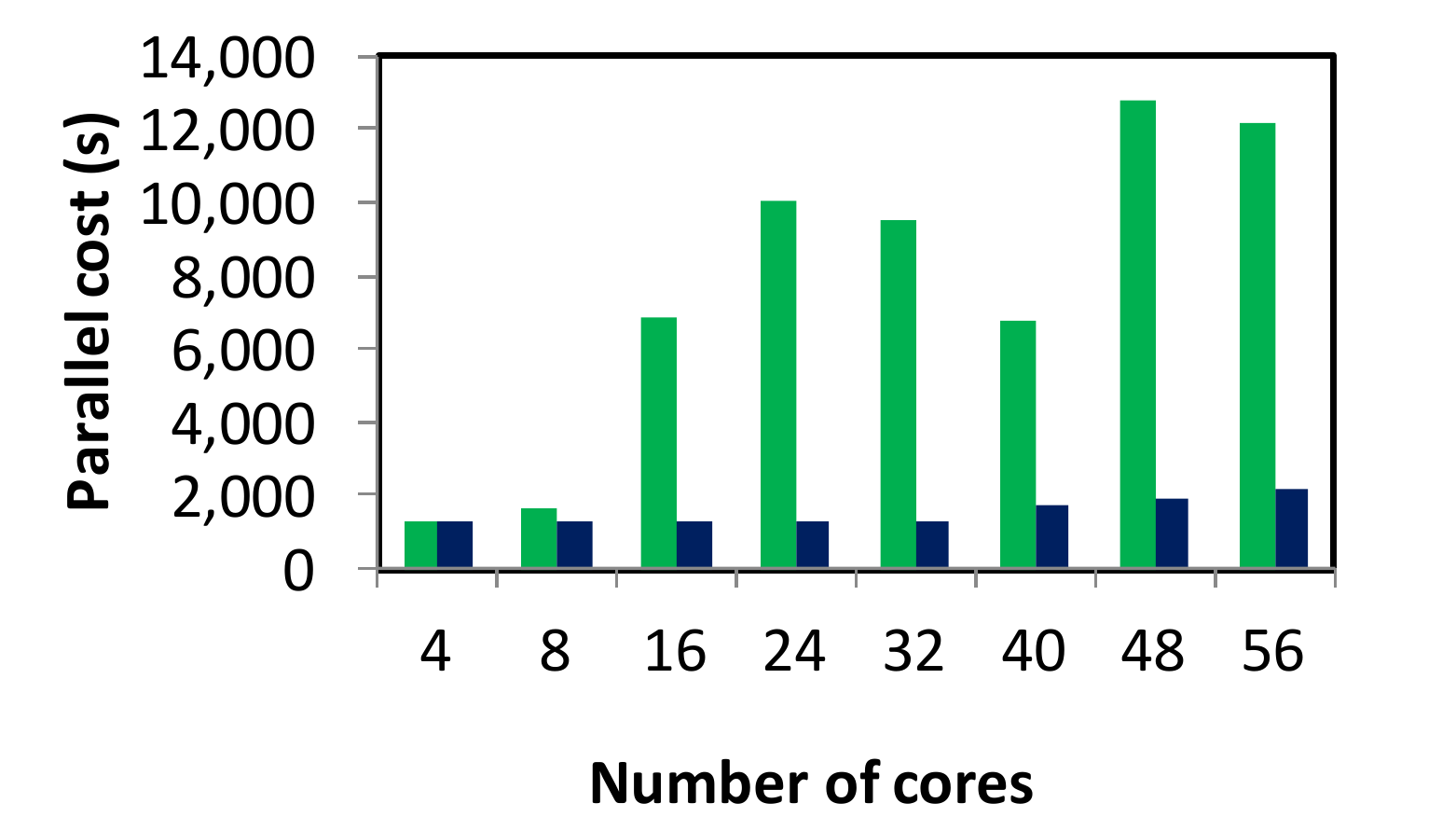}%
		\label{fig3:left3}%
	} 
	\subfloat[FAC \textemdash{} MM]{%
		\includegraphics[scale=0.3, clip,trim=0cm 10cm 18cm 0cm]{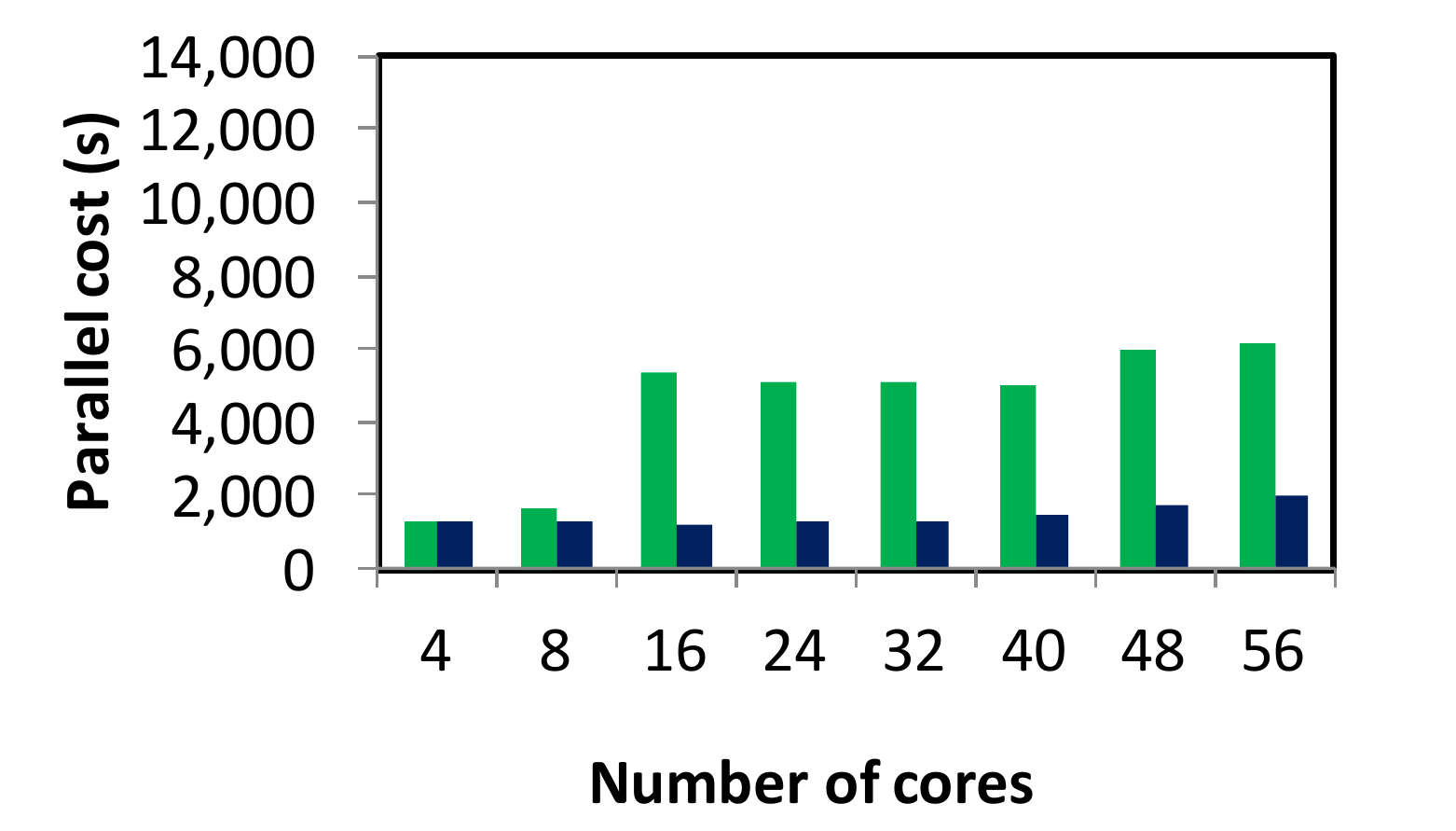}%
		\label{fig3:left4}%
	}
	\\ 
	\subfloat[STATIC \textemdash{} AC-d]{%
		\includegraphics[scale=0.3, clip,trim=0cm 10cm  18cm 0cm]{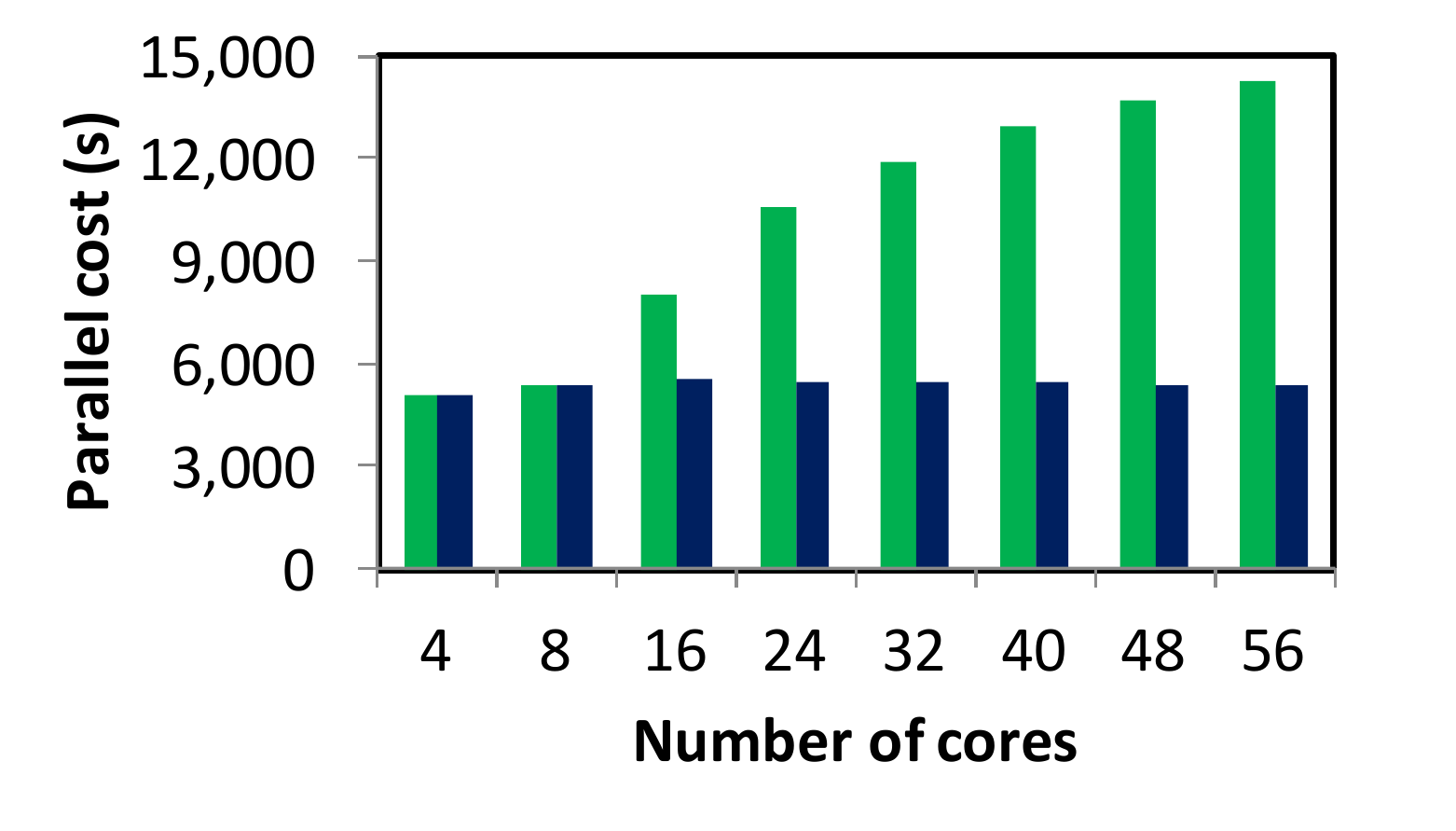}%
		\label{fig3:middle}%
	}
	\subfloat[SS \textemdash{} AC-d]{%
		\includegraphics[scale=0.3, clip,trim=0cm 10cm  18cm 0cm]{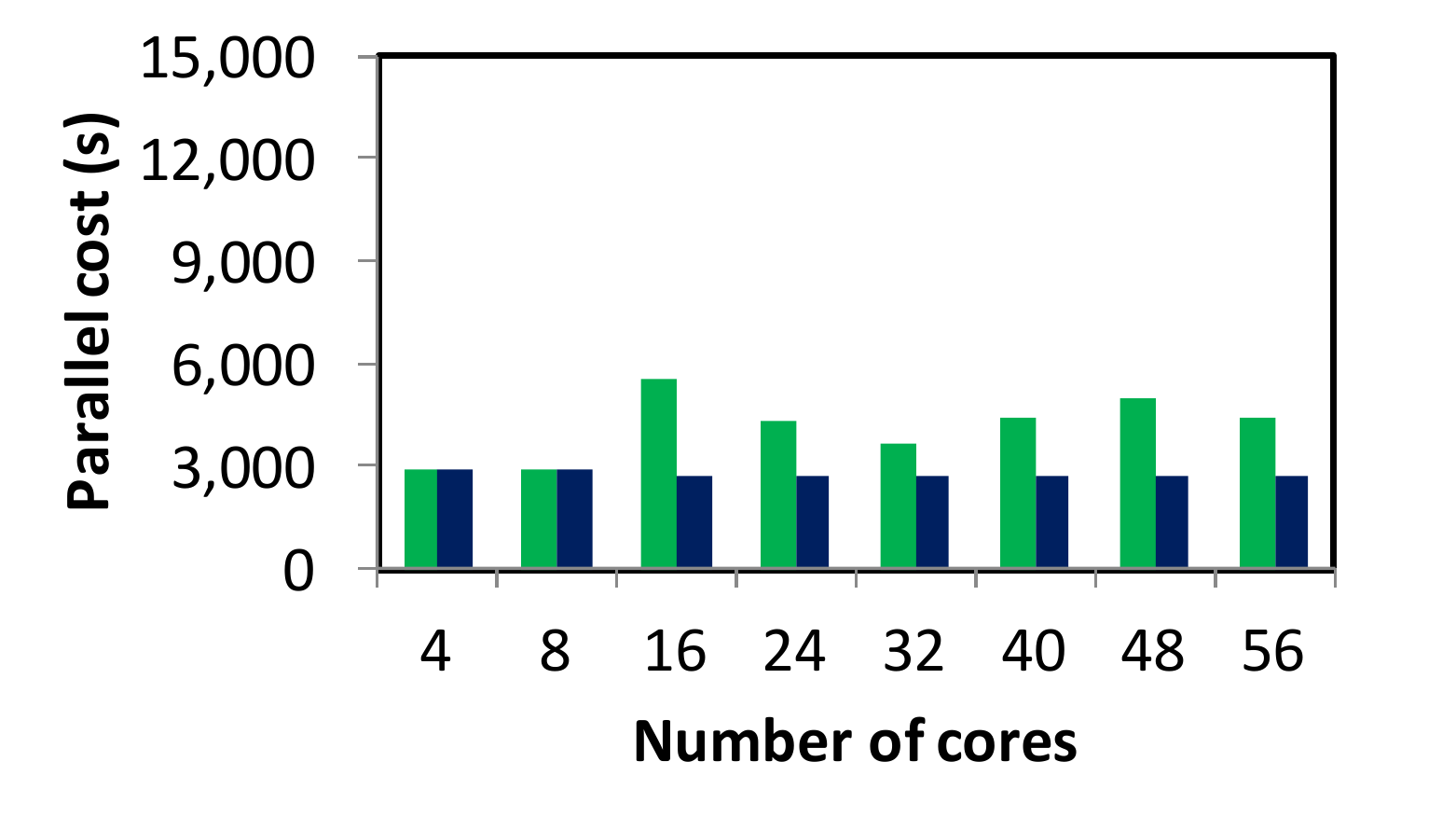}%
		\label{fig3:middle2}%
	}
	\subfloat[GSS \textemdash{} AC-d]{%
		\includegraphics[scale=0.3, clip,trim=0cm 10cm  18cm 0cm]{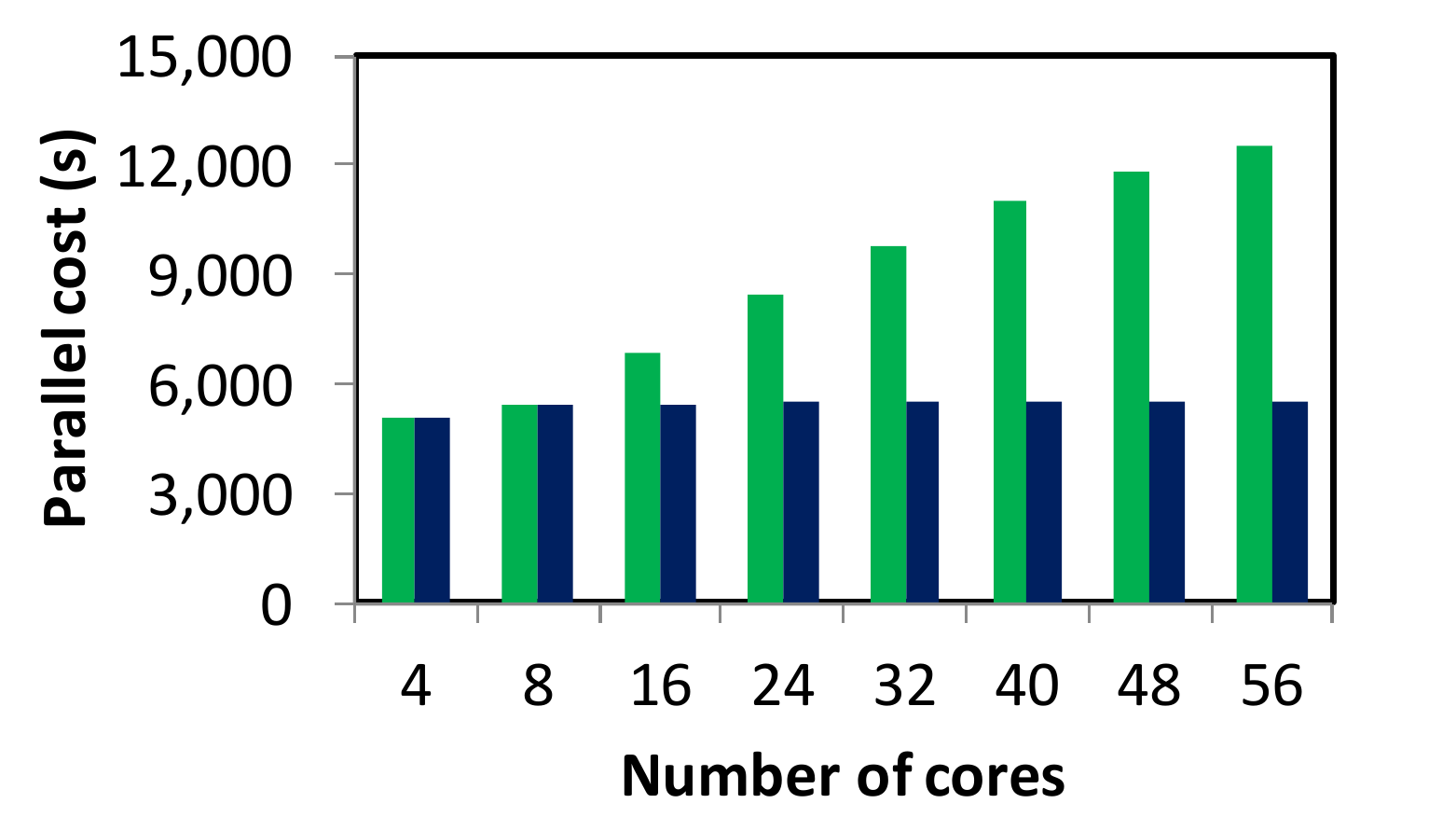}%
		\label{fig3:middle3}%
	}
	\subfloat[FAC \textemdash{} AC-d]{%
		\includegraphics[scale=0.3, clip,trim=0cm 10cm  18cm 0cm]{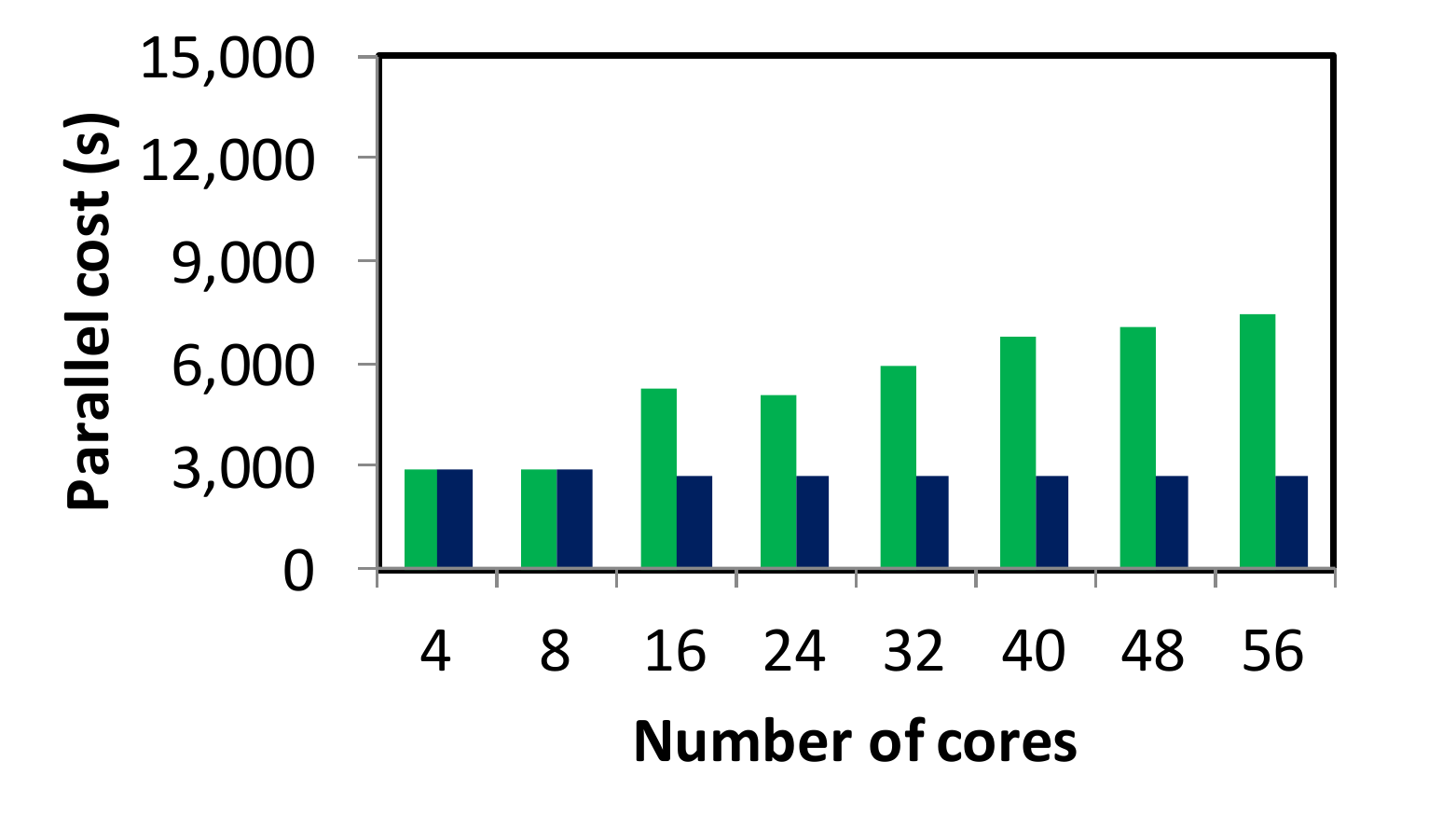}%
		\label{fig3:middle4}%
	}\\
	\subfloat{%
	\includegraphics[scale=0.50, clip,trim=0cm 16.5cm 0cm 1cm]{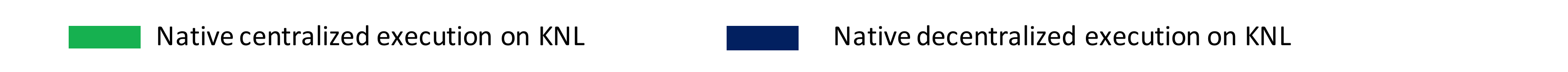}%
	\label{fig3:legend}%
}
	\end{adjustbox}
	\caption{Parallel cost of \alia{the} native execution of selected DLS experiments on the KNL processor with DLS implemented using a decentralized process coordination (dark blue bars) compared with \alia{the} native execution results with DLS implemented using a centralized (master-worker) process coordination (green bars). \mbox{Parallel~cost~=~parallel~program~execution~time $\times$ number~of~threads.}}
	\label{figKNLKNL}
\end{figure*}  
\end{landscape}

\paragraph*{Reproduction Results}
The performance results in terms of parallel cost of the native execution of the DLS techniques implemented using decentralized process coordination are compared to the performance results of the native execution using the centralized approach~\cite{HPCC_FAC} in \figurename{~\ref{figKNLKNL}}. As can be observed from \figurename{~\ref{figKNLKNL}}, the parallel cost of the centralized process coordination increases as the number of threads increases. This can be attributed to the effect of multiple threads competing to lock and unlock the work queue and the master serving work requests and update the same data structure concurrently. In the present decentralized  implementation, threads obtain work on their own from a pool of tasks (accessed via a shared loop index) whenever they become available. The updates of the shared variables between threads are performed using atomic operations. Therefore, the decentralized DLS implementation does not exhibit the same parallel cost increase as the centralized implementation~\cite{HPCC_FAC} as the number of threads increases.
The performance behavior of the four loop scheduling techniques is compared in the execution of the selected experiments on the KNL processor, illustrated in \figurename{~\ref{figKNLKNL}}, and the execution on the RP3 system, illustrated in \figurename{~\ref{figRP3_diff}}.
The performance \alia{trend} of the scheduling techniques in \figurename{~\ref{figKNLKNL}} is \alia{comparable} to that in \figurename{~\ref{figRP3_diff}} for the AC-d kernel, as can be observed by comparing sub-figures (e,f,g,h) in both \figurename{~\ref{figRP3_diff}} and \figurename{~\ref{figKNLKNL}}, \alia{yet the absolute performance is different.} 
However, for the MM kernel, the performance \alia{trend and absolute values} of the scheduling techniques in the past and in the present differs significantly. 
The original results, in \figurename{~\ref{figRP3_diff}}~(b) and~(d), suggest that SS and FAC yield almost similar performance. Examining the results in \figurename{~\ref{figKNLKNL}}~(b) and (d), one can notice a significantly different performance from the one in \figurename{~\ref{figRP3_diff}}~(b) for the same computational kernel. 
In the present results, STATIC, GSS, and FAC outperform SS, which exhibits the poorest performance for the MM kernel. 
The poor performance of SS is due to its large scheduling overhead and the fine granularity of the MM loop iteration for the matrix size under study. 
\alia{The achieved trust in the implementation of DLS techniques for shared-memory systems has already been transferred to their implementation for distributed-memory systems~\cite{Mohammed:2018a}.}
These experiments are performed on a single KNL processor running CentOS operating system version \mbox{3.10.0-327.el7.x86\_64}, with a single 64-core Intel Xeon Phi standalone processor, version~7210. 
The computational kernels’ codes are compiled using GNU C compiler version 6.3.0, with the \mbox{-O3} \mbox{-mavx512f} \mbox{-mavx512cd} \mbox{-mavx512er} \mbox{-mavx512pf} optimization  flags.
 The KNL processor is booted in the memory mode~\cite{KNLmem}. 
 As the computational kernels under study are not memory intensive and the main focus of this work is on studying the effect of scheduling on the execution time, the MCDRAM is, however, not used. 
 All memory allocations being performed on the regular DDR4 memory. 
 To make the results in this work reproducible, the GNU C compiler is used instead of the Intel C compiler. 
 Certain compilation  lags (see above) are used to optimize the generated code for the KNL processor and obtain the best possible performance using the GNU C compiler.

%% file: 7.tex
\section{Prediction of DLS Performance via Simulation}
\label{sec:repKNLsim}
The \simdag{} simulator from Section~\ref{sec:verification} is used to predict the performance of the DLS experiments of interest on the KNL processor using simulation. A \ali{close agreement between the performance prediction using simulation and the native execution} represents an experimental validation of the simulation. \ali{The experimental validation of the simulation is essential to attain trustworthiness in the prediction results of the simulation in future experiments.} 
Accurate parameters that describe the execution on the KNL processor are required as input to the simulator. 
In this work, three such parameters have been identified: 
(1)~Pthreads library: Creation of threads;
(2)~Scheduling overhead;
\ali{and} (3)~Task execution time.\\

Timers are inserted in the source code of the computational kernels around the functions that represent these parameters. 
For instance, timers are inserted before and after Line~2 in Algorithm~\ref{algo:native_code} to measure the scheduling overhead and before and after Line~3 in Algorithm~\ref{algo:native_code} to measure the task execution time.
The measurement procedure described in Section~\ref{sec:repDLSKNL} is used to ensure the accuracy of the time measurements. 
These time overheads (in the microsecond range), are multiplied by the nominal computing speed of a KNL core of $41,600$~MFLOP/s to obtain the computational effort as FLOP. 
The scheduling overheads and tasks execution times are read during simulation from a file.
This way, the simulator can account for these overheads and make more accurate predictions of the execution time.  
As all threads on KNL share the available memory, there is no over~the~network~inter-thread~communication. 
Therefore, the communication size is set to~0~Byte in the \simdag{} simulator for both computational kernels.


\begin{landscape}
\begin{figure*}[b]
	\begin{adjustbox}{minipage=\linewidth,frame}
	\centering
	\subfloat[STATIC \textemdash{} MM]{%
		\includegraphics[scale=0.3, clip,trim=0cm 10cm 18cm 0cm]{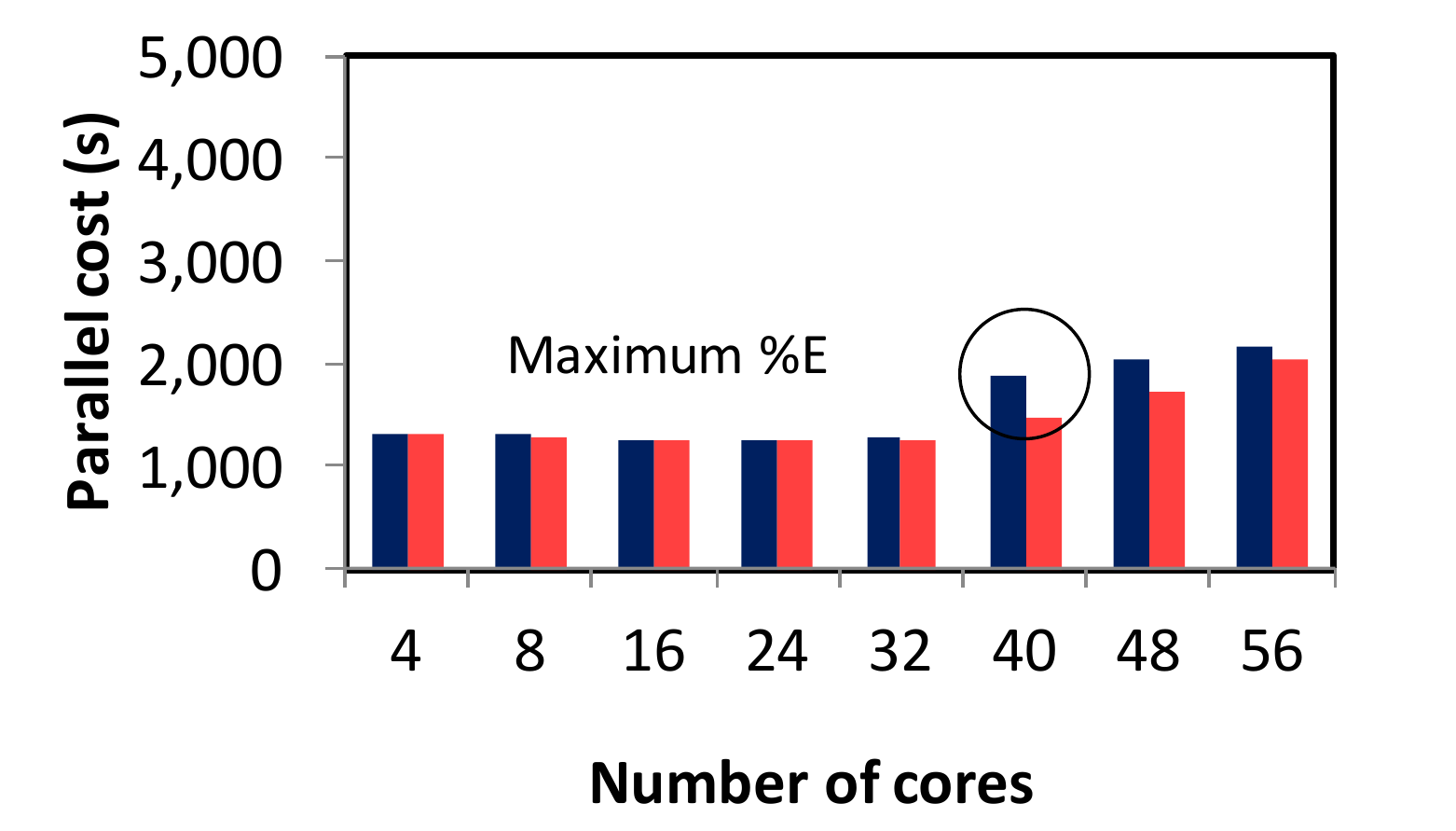}%
		\label{fig4:left}%
	} 
	\subfloat[SS \textemdash{} MM]{%
		\includegraphics[scale=0.3, clip,trim=0cm 10cm 18cm 0cm]{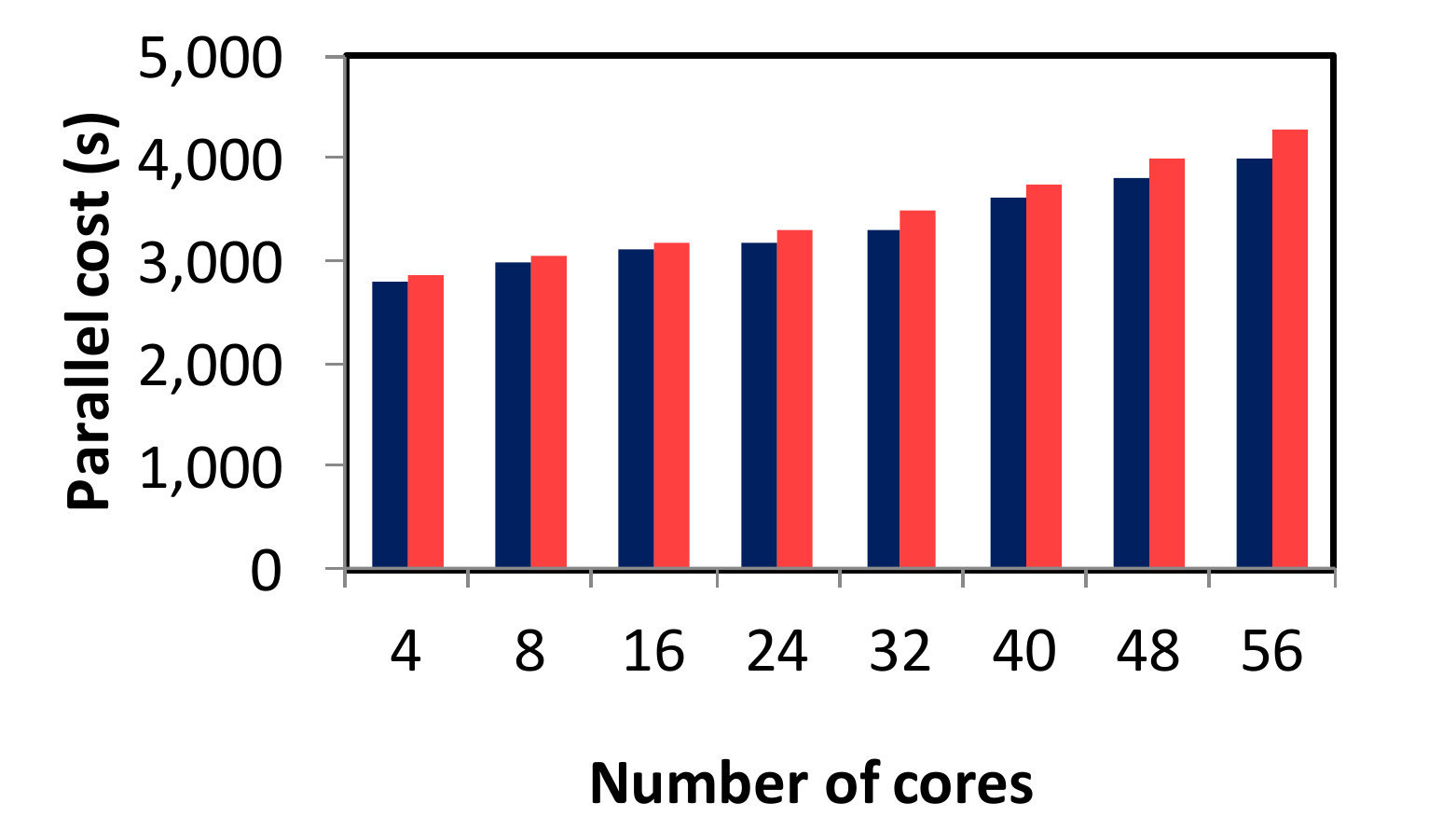}%
		\label{fig4:left2}%
	} 
	\subfloat[GSS \textemdash{} MM]{%
		\includegraphics[scale=0.3, clip,trim=0cm 10cm 18cm 0cm]{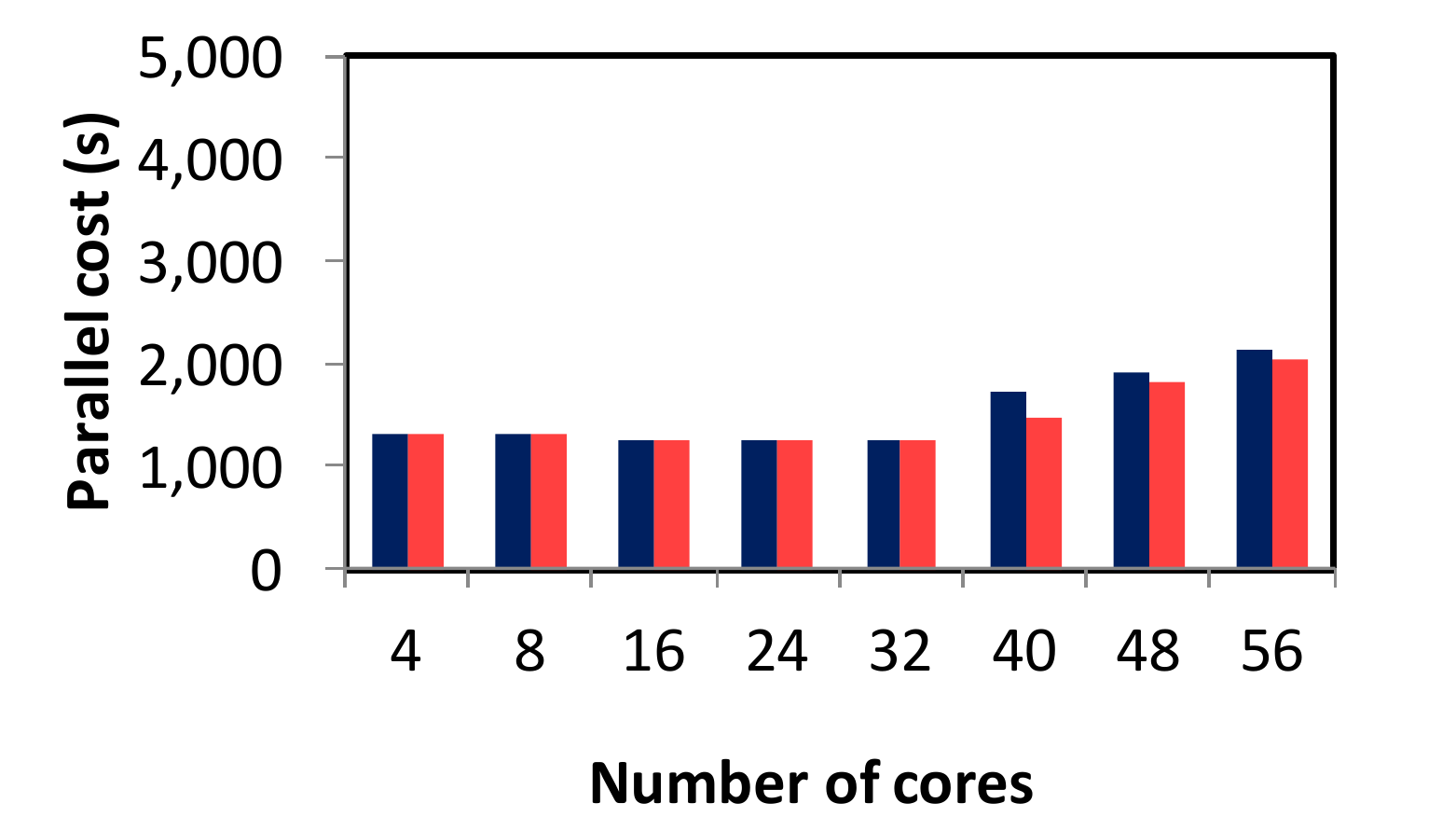}%
		\label{fig4:left3}%
	} 
	\subfloat[FAC \textemdash{} MM]{%
		\includegraphics[scale=0.3, clip,trim=0cm 10cm 18cm 0cm]{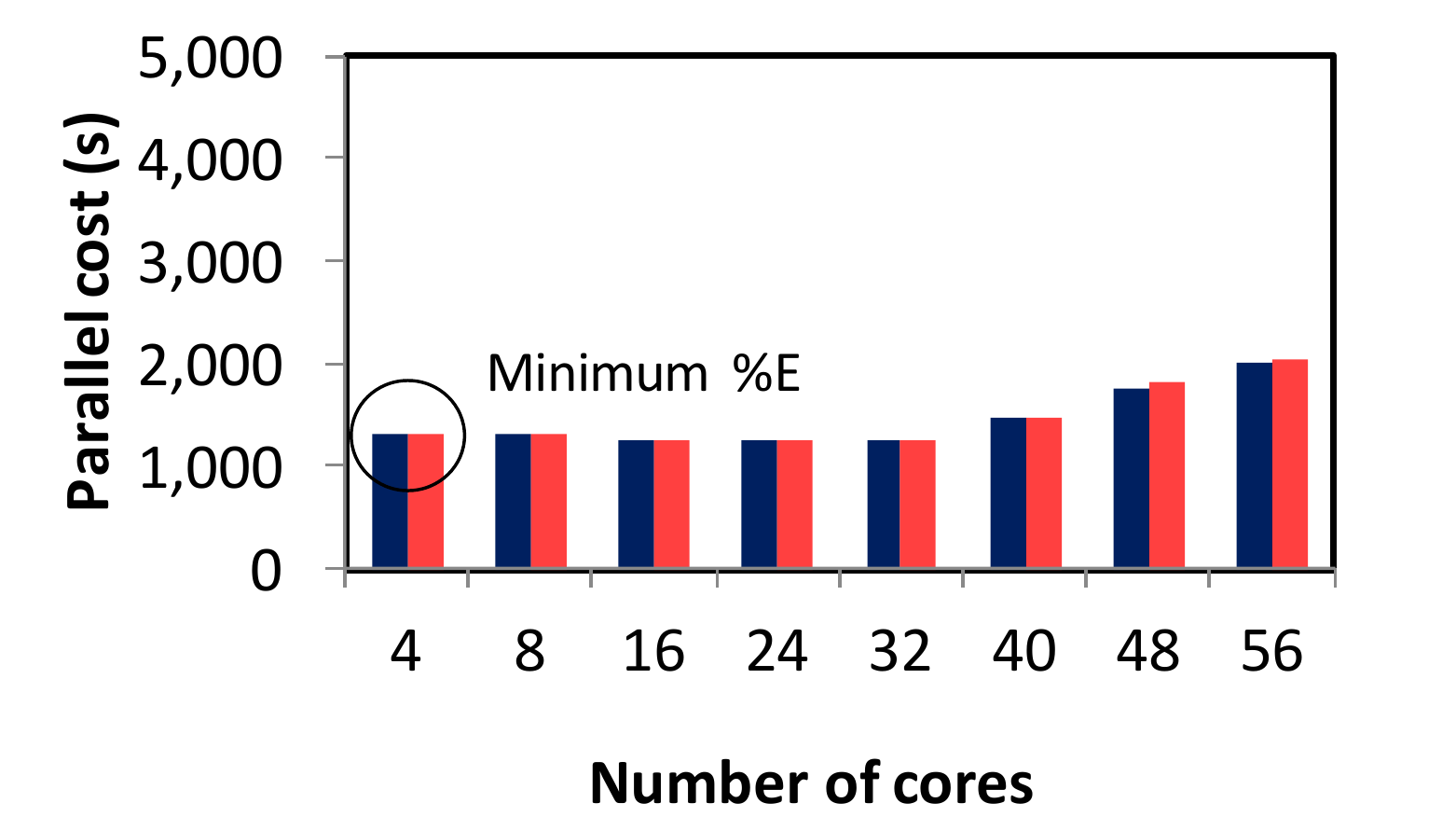}%
		\label{fig4:left4}%
	}
	\\ 
	\subfloat[STATIC \textemdash{} AC-d]{%
		\includegraphics[scale=0.3, clip,trim=0cm 10cm  18cm 0cm]{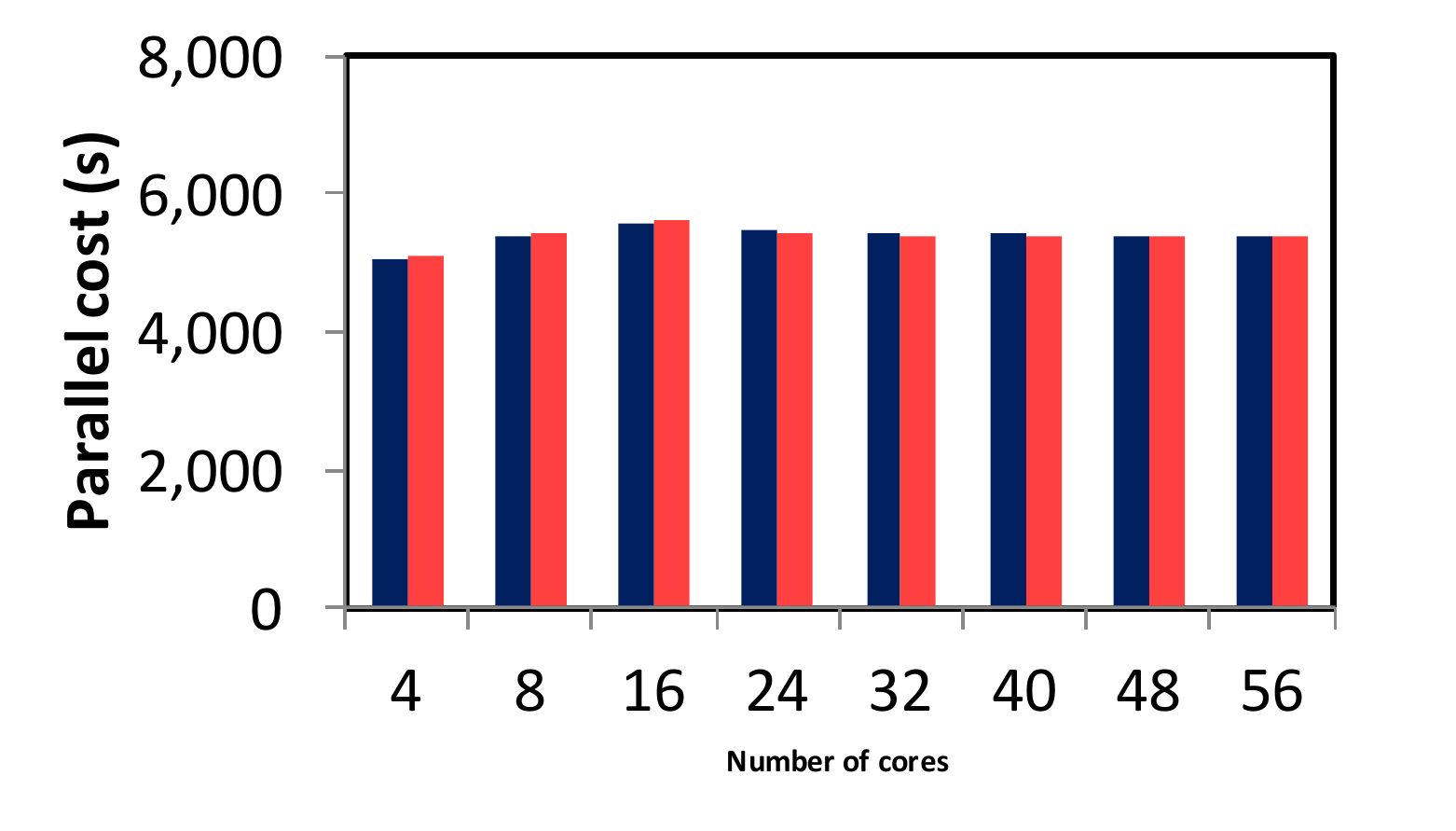}%
		\label{fig4:middle}%
	}
	\subfloat[SS \textemdash{} AC-d]{%
		\includegraphics[scale=0.3, clip,trim=0cm 10cm  18cm 0cm]{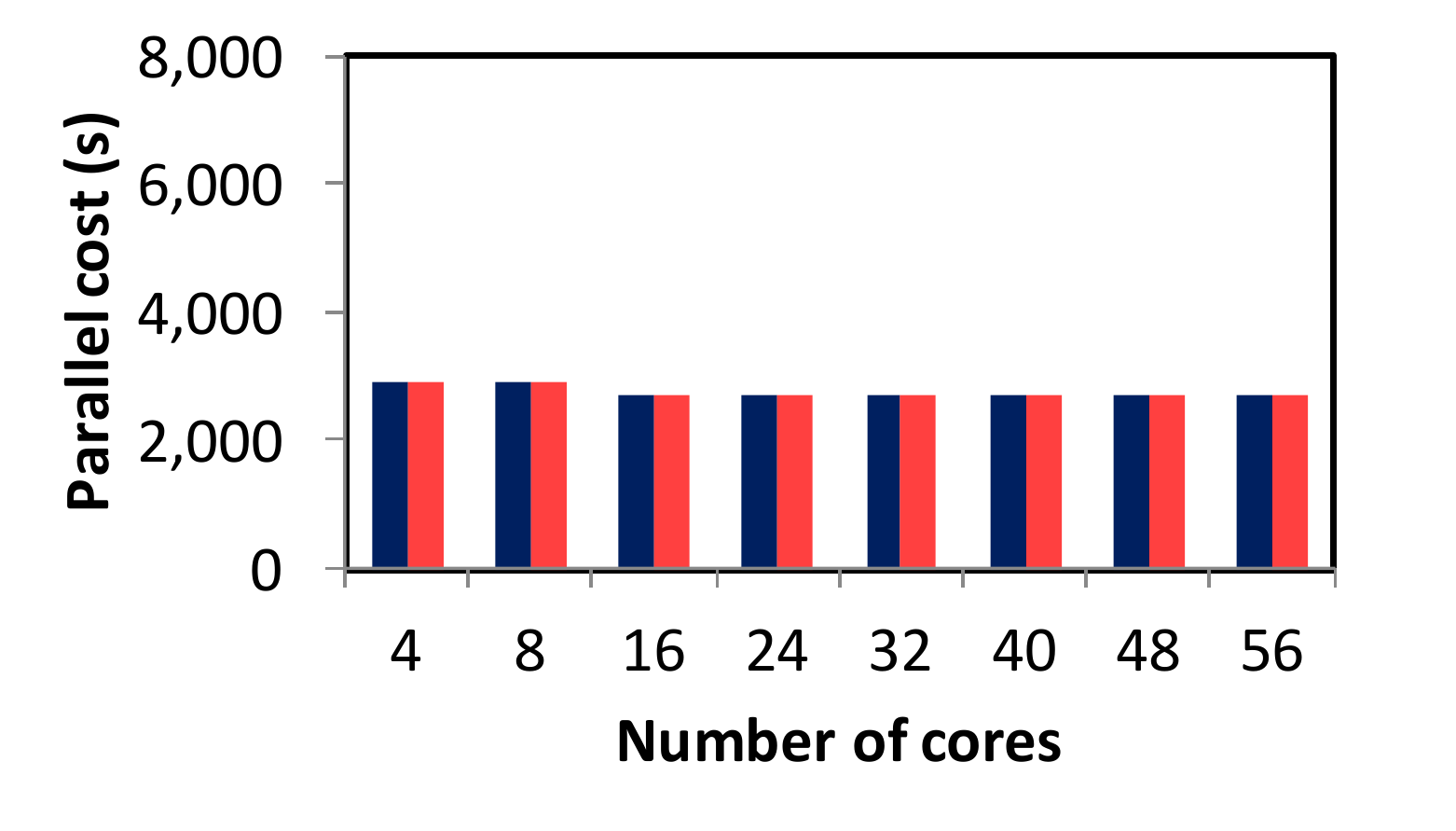}%
		\label{fig4:middle2}%
	}
	\subfloat[GSS \textemdash{} AC-d]{%
		\includegraphics[scale=0.3, clip,trim=0cm 10cm  18cm 0cm]{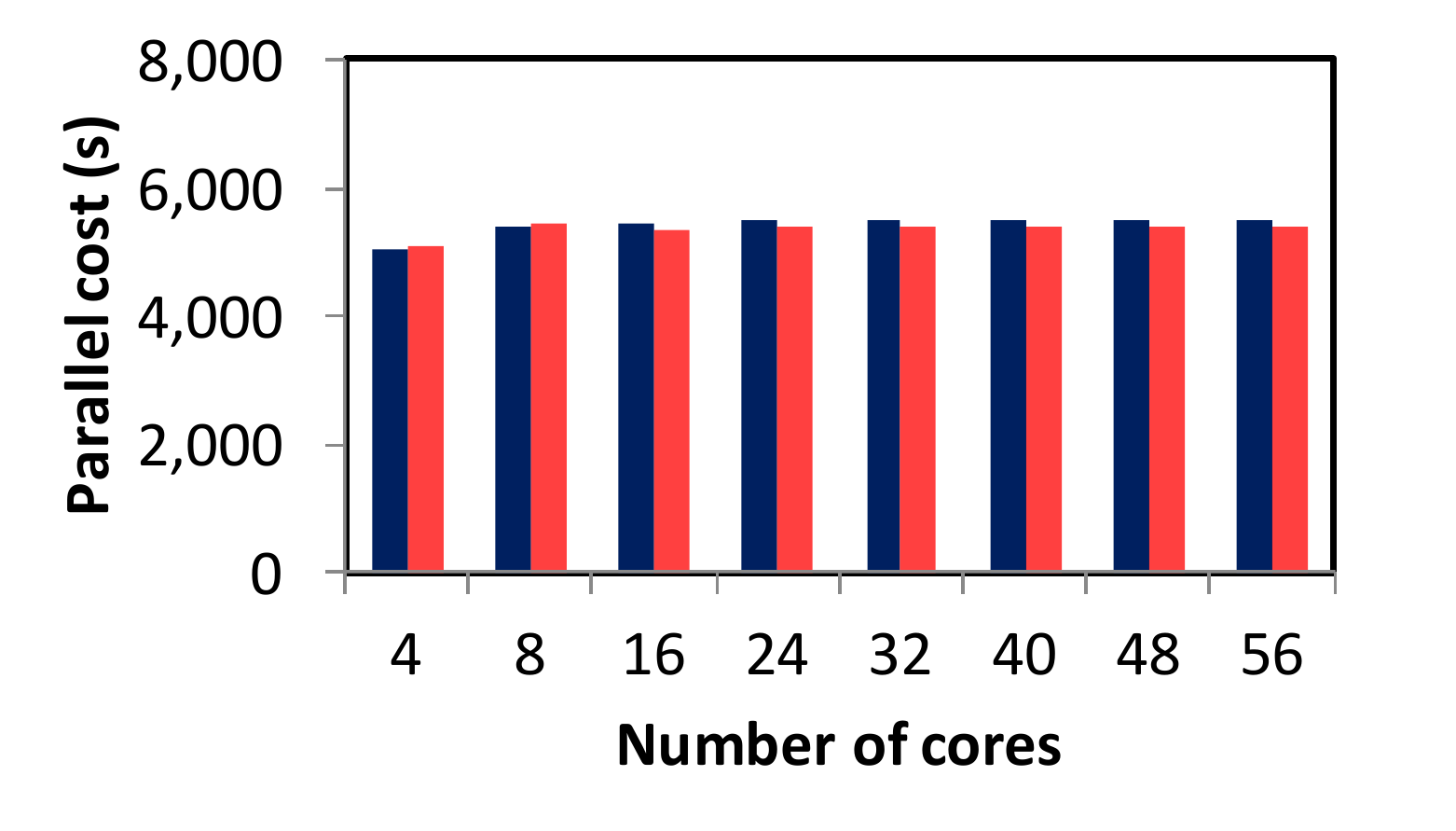}%
		\label{fig4:middle3}%
	}
	\subfloat[FAC \textemdash{} AC-d]{%
		\includegraphics[scale=0.3, clip,trim=0cm 10cm  18cm 0cm]{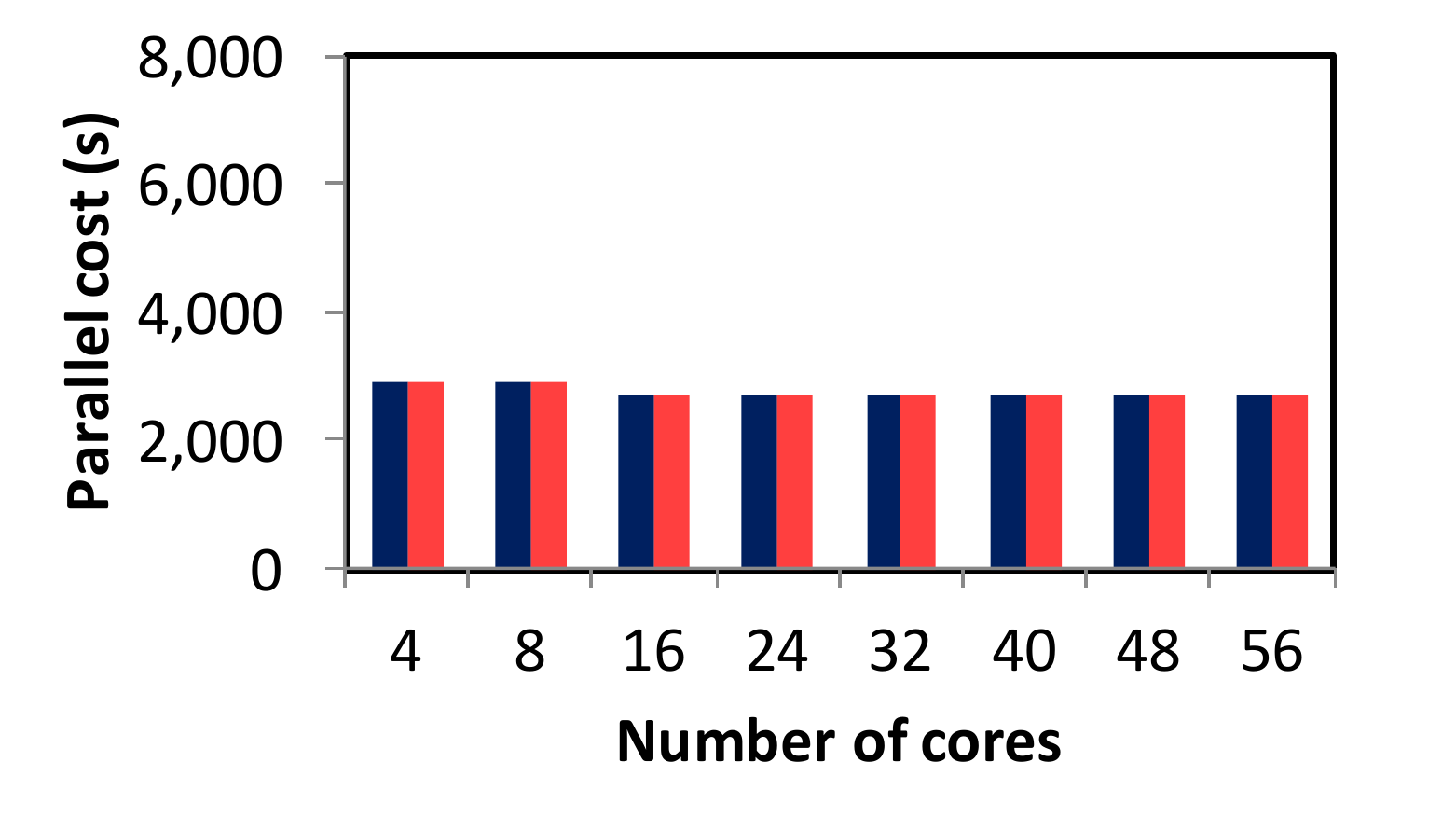}%
		\label{fig4:middle4}%
	}\\
	\subfloat{%
	\includegraphics[scale=0.50, clip,trim=0cm 16.5cm 0cm 1cm]{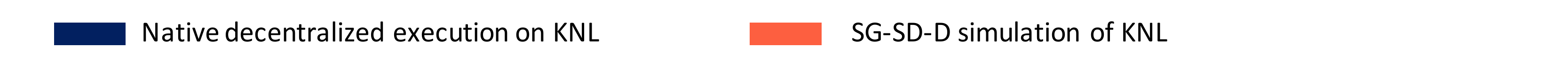}%
	\label{fig4:legend}%
}
\end{adjustbox}
	\caption{\alia{Parallel cost of the execution of selected DLS experiments} obtained with \simdag{}  (red bars) compared with \alia{the cost of the} native execution results on the KNL processor (dark blue bars). \mbox{Parallel~cost~=~parallel~program~execution~time $\times$ number~of~threads.}}
	\label{figKNLdiff}
\end{figure*}    
\end{landscape}
 
The theoretical speed of a single core of the KNL is calculated using information from \ali{the publication that introduced it}~\cite{KNLmem}. 
\ali{The single core speed of the KNL is found to be 41.6~GFLOP/s.}
Even though the \simdag{} simulator, in this experiment, simulates a shared memory system and the network is unused, the parameters for describing the network are still required in the \emph{platform file}. The values that describe the processor speed, network bandwidth, and network latency of the KNL system used in these experiments are $41,600$~MFLOP/s, $100$~Gbit/s, and $100$~ns, respectively.
SimGrid is a multithreaded simulator. To reduce the simulation time of \mbox{SimGrid-based} experiments, it is executed in parallel on the KNL hyperthreaded processor, with 256 hardware threads.
%

\paragraph*{Performance Prediction Results}
The results of the execution of the selected DLS experiments on the KNL are compared to the \mbox{simulation-based} prediction results obtained with \simdag{}-D in \figurename{~\ref{figKNLdiff}}. 
The results show that the simulated execution behavior is in agreement with the native execution for the four different DLS techniques in executing the two computational kernels under \alia{test}. 

The \ali{percent error} $\%E$ between the results of the simulation of KNL execution and the results of KNL execution is calculated as described in Section~\ref{sec:verification}. 
The minimum absolute $\%E$ is $0.00948\%$, for \mbox{FAC \textemdash{} MM} with 4 threads 
as can be observed from \figurename{~\ref{figKNLdiff}}(d). 
The maximum absolute $\%E$ is $21.42\%$, for \mbox{STATIC \textemdash{} MM} with 40 threads
as can be observed from \figurename{~\ref{figKNLdiff}}(a). 
The average of the absolute $\%E$ is $1.94\%$ between all the scheduling experiments on the KNL and their corresponding \simdag{} simulation presented in this work.
The \emph{average percent error} values correspond to acceptable differences. 
More importantly, the performance trends of the studied DLS techniques are similar between the native and simulative executions. 
\textbf{Therefore, it can be stated that the simulator predicts the performance of the two computational kernels with the four implemented scheduling techniques on the present computing system.}
%

%

%% file: 8.tex
\section{Conclusion and Future Work}
\label{sec:conc}

\flo{
In this work, \alia{the} reproduction \ali{of the behavior of two computational kernels~\cite{FAC}} has been used to confirm \alia{the adherence of} the present implementation of four scheduling techniques \alia{to the original specification~\cite{FAC}}.
\alia{The achieved trust in the implementation of DLS techniques for shared-memory systems has also been transferred to their implementation for distributed-memory systems~\cite{Mohammed:2018a}.}
Moreover, the reproduction, in the present, of previous scheduling experiments on modern hardware, is used to evaluate \alia{the hypothesis that} the results and conclusions from past experiments are influenced by the modern software stack and hardware systems used in the present work. 
In contrast to the earlier results of Flynn Hummel {\em et al.}~\cite{FAC} \alia{which} indicate that both FAC and SS perform comparably, this work shows that it is significantly inefficient to use the SS technique for the particular form of the MM kernel considered herein. 
This behavior can be attributed to 
the massive increase in the hardware \ali{computing} speed since 1992~\cite{FAC}, and 
to the fact that \ali{the} MM loop iterations that were considered of large granularity \ali{in earlier work}, are now shown to be of small granularity. 
Consequently, the overhead of allocating work in SS is larger than the time to execute an MM loop iteration. 
Hence, the performance of the SS technique is presently dominated by the scheduling overhead.}
\flo{
The main contribution of this work is a confirmation of the hypothesis that reproducing experiments of identical scheduling scenarios on past and modern hardware may lead to an entirely different behavior from what is expected.}

\flo{A comprehensive study of the performance behavior of dynamic loop scheduling techniques for various applications and on several architectures is envisioned as part of future work.} 
\alia{This work lays the foundation and motivates the reproduction and experimental verification of other DLS techniques and their implementations in other simulators. 
The study of the performance of scientific applications with various DLS techniques under perturbations and failures in the computing system is envisioned as future work.}

%% file: 9.tex
\section*{Acknowledgment} 
This work is partly funded by the Swiss National Science Foundation in the context of the ``Multi-level Scheduling in Large Scale High Performance Computers'' (MLS) grant, number 169123.